\PassOptionsToPackage{colorlinks=true,linkcolor=blue,filecolor=black,urlcolor=blue,citecolor=magenta}{hyperref}
\documentclass[11pt]{article}
\usepackage{jheppub}
\usepackage{amsmath,amssymb}
\usepackage{mathrsfs}
\usepackage{amsfonts}
\usepackage{amsthm}
\usepackage[table]{xcolor}
\usepackage{caption,subcaption}
\usepackage{tikz}
\usetikzlibrary{decorations.markings}
\usetikzlibrary{patterns}
\usepackage{color, soul}
\usepackage{feynmp-auto}
\usepackage{slashed}
\usepackage{arydshln}

\newcommand{\Log}[1]{\log \left( #1 \right)}
\newcommand{\Li}[1]{\text{Li}_2\left( #1 \right)}

\numberwithin{equation}{section}

\newcommand{\eps}{\epsilon}
\preprint{NORDITA 2023-010\\ \vspace{1pt} \hfill UUITP-03/23}
\title{ 
	Inelastic Exponentiation and Classical Gravitational Scattering at One Loop
	}
\author[a]{Alessandro Georgoudis}
\author[b,a]{Carlo Heissenberg}
\author[b,a]{Ingrid Vazquez-Holm}

\affiliation[a]{
	NORDITA, Stockholm University and KTH Royal Institute of Technology,\\Hannes Alfv\'ens v\"ag 12, SE-106 91 Stockholm, Sweden}
\affiliation[b]{Department of Physics and Astronomy, Uppsala University,\\Box 516, 75120 Uppsala, Sweden
}

\abstract{
We calculate the inelastic $2\to3$ one-loop amplitude for the scattering of two point-like, spinless objects with generic masses involving the additional emission of a single graviton. We focus on the near-forward, or classical, limit. Our results include the leading and subleading orders in the soft-region expansion, which captures all non-analytic contributions in the transferred momentum and in the graviton's frequency. This allows us to check the first constraint arising from the inelastic exponentiation put forward in Refs.~\cite{Damgaard:2021ipf,Cristofoli:2021jas,DiVecchia:2022piu}, 
and to calculate the $2\to3$ one-loop matrix element of the $N$-operator, linked to the $S$-matrix by $S = e^{iN}$, showing that it is real, classical and free of infrared divergences.
We discuss how our results feature in the calculation of the $\mathcal O(G^3)$ corrections to the asymptotic waveform.
}

\begin{document}

\maketitle


\section{Introduction and Summary of Results}
\label{sec:introduction}

Recent years have witnessed renewed efforts in the study of two-body systems undergoing classical gravitational collisions, motivated by the ultimate objective of providing increasingly accurate waveform templates for gravitational wave detection \cite{LIGOScientific:2021djp,Buonanno:2022pgc}.
While at first sight counterintuitive, scattering-amplitude methods borrowed from collider physics \cite{Kawai:1985xq,Bern:1994zx,Bern:1994cg,Bern:1995db,Bern:1998ug,Bern:2008qj,Bern:2010ue} have proven to be powerful tools for describing such systems and providing state-of-the-art predictions in the Post-Minkowskian (PM) regime, when the two colliding objects are sufficiently far apart and interact weakly \cite{Bjerrum-Bohr:2018xdl,Cheung:2018wkq,Bern:2019nnu,Bern:2019crd,Bern:2021dqo,Bern:2021yeh}. 
Interactions between astrophysical black holes or neutron stars involved in such collisions are indeed classical, since their typical quantum wavelength is much smaller than the length scale associated to the gravitational curvature they induce, a statement that for black holes of mass $M$ translates to $GM^2/\hbar\gg1$ \cite{Bern:2019crd,DiVecchia:2021ndb,Bellazzini:2022wzv}. This inequality, which is of course amply satisfied by such objects, signals, however, a breakdown of conventional perturbation theory, since the effective coupling to gravity is not small. Therefore scattering amplitudes, which are organized as a weak-coupling $G$-expansion, must  actually be supplemented with a nonperturbative principle in order to correctly capture the classical limit. 

One such guiding principle, which is also familiar from the non-relativistic WKB approximation, is that in the classical limit the $S$-matrix should be dominated by the exponential $e^{2i\delta}$ of a large phase $2\delta$, which plays the role of a large action in units of $\hbar$. For the elastic $2\to2$ amplitude this resummation is known as the eikonal exponentiation and the eikonal phase, or the closely related radial action, has been employed to extract from the amplitude the deflection angle(s) for two-body collisions up to 4PM order \cite{Kabat:1992tb,Akhoury:2013yua,KoemansCollado:2019ggb,Cheung:2020gyp,DiVecchia:2020ymx,AccettulliHuber:2020oou,DiVecchia:2021ndb,DiVecchia:2021bdo,Brandhuber:2021kpo,Heissenberg:2021tzo,Bjerrum-Bohr:2021vuf,Bjerrum-Bohr:2021din,Brandhuber:2021eyq,DiVecchia:2022nna,Bern:2021dqo,Bern:2021yeh}.
The nonperturbative nature of the problem manifests itself at each loop order via ``superclassical'' or ``iteration'' terms, contributions that scale with higher powers of the large ratio $GM^2/\hbar$, or, for short, of $\hbar^{-1}$. The eikonal exponentiation dictates how such spurious terms should be subtracted, by matching with the power series expansion of the exponential, and fixes all ambiguities associated to possible remainders, providing a direct connection to the impulse and to the deflection angle via a saddle point approximation \cite{Amati:1987uf,Amati:1987wq,Amati:1990xe,Amati:1992zb,Amati:1993tb,Ciafaloni:2014esa,DiVecchia:2021bdo}. 

However, by focusing on the elastic $2\to2$ amplitude, the conventional eikonal framework fails to capture possible subtractions associated to inelastic channels. For instance the infrared (IR) divergent imaginary part in the 3PM eikonal signals the fact that at $\mathcal O(G^3)$ an inelastic 3-particle channel involving the two massive states and a graviton opens up, and the standard eikonal exponentiation does not capture it.
This problem has been studied and solved in \cite{Damgaard:2021ipf,DiVecchia:2022nna,Cristofoli:2021jas,DiVecchia:2022piu}, the basic idea being that, in a more comprehensive framework, the eikonal should be promoted to the exponential of $i$ times a suitable Hermitian operator that is able to appropriately combine all needed channels. The approach of Ref.~\cite{Damgaard:2021ipf} is to apply this principle to the full $S$-matrix, writing $S = e^{iN}$ with $N^\dagger= N$ and building $N$-matrix elements out of  conventional scattering amplitudes, which are of course $T$-matrix elements with $S = 1 + iT$.

A complementary approach is provided by the formalism first introduced by Kosower, Maybee and O'Connell (KMOC) \cite{Kosower:2018adc} and later developed in Refs.~\cite{Herrmann:2021lqe,Herrmann:2021tct,Cristofoli:2021vyo,Cristofoli:2021jas,Britto:2021pud,Adamo:2022rmp,Adamo:2022qci}. This framework is based on the principle that, after identifying a well-defined classical observable $O$ associated to the collision, its expectation value in the final state dictated by the $S$-matrix, $\langle \text{in}| S^\dagger O S |\text{in}\rangle$ 
will be free of superclassical terms and thus possesses a well-defined classical limit. The state $|\text{in}\rangle$ models the two incoming massive particles with given impact parameter(s) via an appropriate superposition of plane-wave states built with suitable wave-packets, whose details become immaterial after the cancellation of superclassical terms.

In this paper, we explore further the exponentiation in the classical limit and the connection between amplitudes and classical observables. We focus on the $2\to3$ amplitude for the scattering of two minimally-coupled massive scalars plus the emission of a single graviton in General Relativity, whose loop expansion reads $\mathcal{A}^{\mu\nu}=\mathcal{A}_0^{\mu\nu}+\mathcal{A}_1^{\mu\nu}+\cdots$, or, pictorially,
\begin{equation}
	\begin{gathered}
\begin{tikzpicture}
	\path [draw, ultra thick, blue] (-4.5,2.2)--(-1.5,2.2);
	\path [draw, ultra thick, green!60!black] (-4.5,.8)--(-1.5,.8);
	\path [draw, red] (-3,1.5)--(-1.5,1.5);
	\filldraw[white, very thick] (-3,1.5) ellipse (.7 and 1);
	\filldraw[pattern=north west lines, very thick] (-3,1.5) ellipse (.7 and 1);
\end{tikzpicture}
\end{gathered}\,
=
\,
\begin{gathered}
	\begin{tikzpicture}
		\path [draw, ultra thick, blue] (-4.5,2.2)--(-1.5,2.2);
		\path [draw, ultra thick, green!60!black] (-4.5,.8)--(-1.5,.8);
		\path [draw, red] (-3,1.5)--(-1.5,1.5);
		\filldraw[black!20!white, thick] (-3,1.5) ellipse (.7 and 1);
		\draw[thick] (-3,1.5) ellipse (.7 and 1);
	\end{tikzpicture}
\end{gathered}
\,
+
\,
\begin{gathered}
	\begin{tikzpicture}
		\path [draw, ultra thick, blue] (-4.5,2.2)--(-1.5,2.2);
		\path [draw, ultra thick, green!60!black] (-4.5,.8)--(-1.5,.8);
		\path [draw, red] (-3,1.5)--(-1.5,1.5);
		\filldraw[black!20!white, thick] (-3,1.5) ellipse (.7 and 1);
		\draw[thick] (-3,1.5) ellipse (.7 and 1);
		\filldraw[white, thick] (-3,1.5) ellipse (.3 and .4);
		\filldraw[pattern=none, thick] (-3,1.5) ellipse (.3 and .4);
	\end{tikzpicture}
\end{gathered}
\,
+\cdots
\end{equation}
and discuss the calculation of its 1-loop part, $\mathcal{A}_1^{\mu\nu}$, starting from the integrand provided in Ref.~\cite{Carrasco:2021bmu}. We focus on the non-analytic terms in the near-forward limit, whereby the transferred momentum and the emitted graviton's momentum are simultaneously taken to be small, $\mathcal O (\hbar)$, in comparison with the particles' masses, $\mathcal O (\hbar^0)$. To this end we apply the method of regions and restrict our attention to the soft region, in which the loop momentum assigned to the exchanged gravitons is also small, $\mathcal O (\hbar)$. We employ dimensional regularization, letting $\epsilon= \frac{4-D}{2}$, and express the result as a Laurent expansion around $\epsilon=0$.

The calculation of $\mathcal{A}_1^{\mu\nu}$ in the soft region constitutes one of the main new results of this work, and represents a first step in generalizing the studies of graviton emissions during collisions of ultrarelativistic or massless objects \cite{Ciafaloni:2015xsr,Ciafaloni:2018uwe,Addazi:2019mjh} to the case of massive objects with generic velocities. The amplitude $\mathcal A_1^{\mu\nu}$, as expected, involves both superclassical, $\mathcal O (\hbar^{-2})$, and classical, $ \mathcal O (\hbar^{-1})$, contributions, for each of which we calculate both infrared (IR) divergent and finite terms. For the IR divergent pieces, we find complete agreement with the well-known exponential pattern \cite{Weinberg:1965nx} according to which IR divergences in a given one-loop amplitude are equal to a one-loop-exact divergent factor $\mathcal W$ times the tree-level amplitude with the same external states \cite{Luna:2017dtq,DiVecchia:2021bdo}.

In the Weinberg limit, in which the emitted graviton's frequency becomes very small, $k^\mu\sim \mathcal O(\lambda)$ with $\lambda\to0$, the one-loop amplitude $\mathcal A_1^{\mu\nu}$ must also exhibit $\mathcal O(\lambda^{-1})$ terms whose form is completely fixed by the leading soft graviton theorem \cite{Weinberg:1964ew} as the factor $\mathcal F^{\mu\nu} = \sqrt{8\pi G}\sum_n p_n^\mu p_n^\nu / (p_n\cdot k)$ times the $2\to2$ one-loop amplitude without graviton emissions. This $1/\lambda$ pole is a frequency-space manifestation of the memory effect \cite{Strominger:2014pwa,Strominger:2017zoo}. Comparing with the results available from the literature \cite{Cheung:2018wkq,KoemansCollado:2019ggb,Cristofoli:2020uzm}, we find perfect agreeement with this prediction, reproducing in particular the terms arising from the 2PM deflection encoded in one-loop $2\to2$ ``triangle'' contributions, i.e.~from the sub-leading eikonal phase $2\delta_1$.
Moreover, exploiting the conventional exponentiation of the $2\to2$ amplitude, this factorization allows us to check the inelastic exponentiation of Refs.~\cite{Cristofoli:2021jas,DiVecchia:2022piu} to leading order in the soft limit. Throughout the paper, we focus on emitted gravitons with \emph{positive} frequencies, so that we do not include in our analysis  terms with support localized at $\omega=0$ in frequency space, which are associated to static effects in time domain (see e.g.~\cite{Mougiakakos:2021ckm,Riva:2021vnj} for their concrete appearance in the tree-level expressions). The inclusion of such terms has been discussed in \cite{Manohar:2022dea,DiVecchia:2022owy} and can be typically performed by means an appropriate dressing of the initial and final states with a modified Weinberg factor $\sqrt{8\pi G}\sum_n p_n^\mu p_n^\nu / (p_n\!\cdot\! k-i0)$.

After constructing the appropriate subtractions dictated by the $N$-matrix formalism \cite{Damgaard:2021ipf}, we calculate the $2\to3$, $N$-matrix element $\mathcal B_1^{\mu\nu}$  from the amplitude $\mathcal A_1^{\mu\nu}$. In this way we obtain a purely classical object, $\mathcal B_1^{\mu\nu}$, which is also real and free of IR divergences. 
Indeed, by comparing the operator power series $N = -i\log(1+iT) = T - \frac{i}{2}T^2+ \cdots$ and the unitarity constraint $i(T^\dagger-T) = T^\dagger T$, it is easy to see that, at one loop, the operator exponentiation of \cite{Damgaard:2021ipf} boils down to simply dropping the imaginary parts of the amplitude, i.e.~to subtracting its unitarity cuts, $2\operatorname{Im}\mathcal A_1^{\mu\nu} = \sum(\text{cuts})$. For the process under considerations, there are four distinct channels, which we depict in Table~\ref{tab:channels}.
\begin{table}
	\begin{center}
	\begin{tabular}{r c c l}
$S$-channel
&
\begin{minipage}{.3\textwidth}
			\begin{tikzpicture}
				\path [draw, ultra thick, blue] (-4,2)--(-.3,2);
				\path [draw, ultra thick, green!60!black] (-4,1)--(-.3,1);
				\path [draw, red] (-1,1.5)--(-.32,1.5);
				\filldraw[black!20!white, thick] (-3,1.5) ellipse (.5 and .8);
				\draw[thick] (-3,1.5) ellipse (.5 and .8);
				\filldraw[black!20!white, thick] (-1.3,1.5) ellipse (.5 and .8);
				\draw[thick] (-1.3,1.5) ellipse (.5 and .8);
			\end{tikzpicture}
\end{minipage}
&
\begin{minipage}{.3\textwidth}
			\begin{tikzpicture}
				\path [draw, ultra thick, blue] (-4,2)--(-.3,2);
				\path [draw, ultra thick, green!60!black] (-4,1)--(-.3,1);
				\path [draw, red] (-3,1.5)--(-2.1,1.5);
				\filldraw[black!20!white, thick] (-3,1.5) ellipse (.5 and .8);
				\draw[thick] (-3,1.5) ellipse (.5 and .8);
				\filldraw[black!20!white, thick] (-1.3,1.5) ellipse (.5 and .8);
				\draw[thick] (-1.3,1.5) ellipse (.5 and .8);
			\end{tikzpicture}
		\end{minipage}
& $\mathcal O(\hbar^{-2})$
	\vspace{.5cm}
		\\
	$C$-channel &
	\begin{minipage}{.3\textwidth}
			\begin{tikzpicture}
				\path [draw, ultra thick, blue] (-4,2)--(-.3,2);
				\path [draw, ultra thick, green!60!black] (-4,1)--(-2.1,1);
				\path [draw, red] (-3,1.5)--(-.3,1.5);
				\filldraw[black!20!white, thick] (-3,1.5) ellipse (.5 and .8);
				\draw[thick] (-3,1.5) ellipse (.5 and .8);
				\filldraw[black!20!white, thick] (-1.3,1.75) ellipse (.45 and .55);
				\draw[thick] (-1.3,1.75) ellipse (.45 and .55);
			\end{tikzpicture}
		\end{minipage}
	&
	\begin{minipage}{.3\textwidth}
			\begin{tikzpicture}
				\path [draw, ultra thick, blue] (-4,2)--(-2.1,2);
				\path [draw, ultra thick, green!60!black] (-4,1)--(-.3,1);
				\path [draw, red] (-3,1.5)--(-.3,1.5);
				\filldraw[black!20!white, thick] (-3,1.5) ellipse (.5 and .8);
				\draw[thick] (-3,1.5) ellipse (.5 and .8);
				\filldraw[black!20!white, thick] (-1.3,1.25) ellipse (.45 and .55);
				\draw[thick] (-1.3,1.25) ellipse (.45 and .55);
			\end{tikzpicture}
		\end{minipage}
& $\mathcal O(\hbar^{-1})$
	\end{tabular}
	\end{center}
\caption{The $s$, $s'$, $s_1$, $s_2$ channels and their scaling in the classical limit. \label{tab:channels}}
\end{table}

Two of them, often referred to as $s$ and $s'$, involve cutting an intermediate state with two massive particles, and we shall call them collectively ``$S$-channel''. We find that the subtraction of the $S$-channel is actually enough to get rid of all superclassical terms. This is in accordance with Refs.~\cite{Cristofoli:2021jas,DiVecchia:2022piu}, since this subtraction in momentum space is equivalent in $b$-space to the subtraction of $2i\delta_0\,\tilde{\mathcal{A}}_0^{\mu\nu}$. Indeed, since this cut contributes schematically via $+\frac{i}{2}(S\text{-channel})$ to the amplitude and each of the two diagrams in the first line of Table~\ref{tab:channels} contributes as $2\delta_0\,\tilde{\mathcal{A}}_0^{\mu\nu}$ in $b$-space, it is crucial to consider both diagrams in order to get the right combinatoric factor in front. 

The remaining two cuts in Table~\ref{tab:channels}, $s_1$ and $s_2$, instead involve an intermediate state with a massive particle and a graviton, in which the latter re-scatters against the massive line in the gravitational analog of a Compton process. For this reason, we may call them collectively ``$C$-channel''.
Albeit classical as far as the $\hbar$ scaling is concerned, the $C$-channel involves an infrared divergence and its subtraction is crucial in order to make the resulting $\mathcal B_1^{\mu\nu}$ (real and) finite as $\epsilon\to0$.

The amplitude $\mathcal A_1^{\mu\nu}$, and the $N$-matrix element $\mathcal B_1^{\mu\nu}$, encode the dynamical information needed  in order to evaluate the $\mathcal O(G^3)$ corrections to asymptotic value of the metric fluctuation far away from the collision, which provide the next order in the PM expansion compared to the results of Refs.~\cite{DEath:1976bbo,Kovacs:1977uw,Kovacs:1978eu,Jakobsen:2021smu,Mougiakakos:2021ckm,Riva:2021vnj}. For this reason we investigate the construction of the associated KMOC kernel, i.e.~the object whose Fourier transform from $q$-space to $b$-space provides the waveform in frequency domain. 
We find that this kernel is not simply given by $i\mathcal B_1^{\mu\nu}$, as perhaps expected. Rather, it equals $i\mathcal B_1^{\mu\nu}$ minus $\tfrac 12$ times the IR divergent $C$-channel cuts. By its very nature, the associated IR pole in $1/\epsilon$ can be exponentiated to a $q$-independent phase, amounting to a (divergent) shift of the origin of the observer's retarded time, at the price of introducing a logarithm involving an unspecified scale in the finite part. Neither the phase nor this logarithm appear in the energy-momentum spectra, and could as such be considered ``harmless''.
The appearance of ambiguous logarithms in the waveform, as a result of the long-range nature of the gravitational force, is a known issue \cite{Goldberger:2009qd,Porto:2012as} and is associated to an ambiguity in the definition of the asymptotic detector's retarded time induced by so-called tail or rescattering effects \cite{Blanchet:1992br,Blanchet:1993ec,Blanchet:2013haa,Bini:2022enm}. We leave further investigations of this issue for future work, together with the calculation of the ($b$-space) waveform and with a comparison with subleading $\log\lambda$-corrected soft theorems \cite{Laddha:2018vbn,Sahoo:2018lxl,Saha:2019tub,Sahoo:2021ctw}, which are also intimately related with long-range corrections to the asyptotic interactions.

The paper is organized as follows.
In Section~\ref{sec:kinematics} we present our conventions for dealing with the external states of the scattering, illustrating a useful choice of variables and polarizations, while more details on them are available in Appendix~\ref{appendix:Kinematics}. In Section~\ref{sec:classicallimit} we discuss the classical limit and how focusing on the soft region  simplifies the integration. We list the corresponding 9 independent master integrals in Appendix~\ref{appendix:Masters} and in the file \texttt{master\_integrals.m}. Section~\ref{sec:structure} is devoted to illustrating our result for the amplitude, which is collected in computer-readable format in the file \texttt{Results\_Ampl\_5pt.nb}, discussing the consistency checks offered by the exponentiation of IR divergences and from factorization in the soft limit. We also discuss the implications of unitarity and the subtraction of superclassical iterations, explicitly proving the leading constraint coming from the inelastic exponentiation. The tree-level amplitudes needed to perform such checks are presented in Appendix~\ref{appendix:treelevel}.
In Section~\ref{sec:waveform}, we discuss the calculation of the gravitational field, of the associated spectrum and of the asymptotic waveform, before presenting a summary of our conclusions and a prospect of possible future directions in Section~\ref{sec:outlook}.

\paragraph*{Conventions:} We employ the mostly-plus signature, $\eta_{\mu\nu} = \operatorname{diag}(-,+,+,+)$. All momenta are regarded as formally outgoing, so that $p_3$, $p_4$ and $k$ are the physical momenta of the final states of the scattering, while $p_1$ and $p_2$ are \emph{minus} the physical momenta of the initial states.

\paragraph*{Note added 1:} While working on this project we became aware of independent progress by 
Refs.~\cite{Brandhuber:2023hhy,Herderschee:2023fxh,Elkhidir:2023dco}, whose scope partly overlaps with our analysis.
These groups' work was also presented as a series of seminars \cite{BrownTalk,RoibanTalk,SergolaTalk} at the ``QCD Meets Gravity 2022'' conference.

\paragraph*{Note added 2:}
In Ref.~\cite{Caron-Huot:2023vxl} it was pointed out that an extra term must be taken into account in the KMOC cuts contributing to the one-loop waveform discussed in Section~\ref{sec:waveform}. See Refs.~\cite{Georgoudis:2023eke,Georgoudis:2023ozp} for its inclusion.

\section{Kinematics}
\label{sec:kinematics}

We consider the scattering of two massive objects with masses $m_1$ (depicted with a thick blue line) and $m_2$ (thick green line) and the emission of a graviton (thin red line),
\begin{equation}\label{Aq1q2k}
	\mathcal{A}^{\mu\nu} = 	\begin{gathered}
		\begin{tikzpicture}[scale=1]
			\draw[<-] (-4.8,5.17)--(-4.2,5.17);
			\draw[<-] (-1,5.15)--(-1.6,5.15);
			\draw[<-] (-1,3.15)--(-1.6,3.15);
			\draw[<-] (-1,.85)--(-1.6,.85);
			\draw[<-] (-4.8,.83)--(-4.2,.83);
			\draw[<-] (-2.85,1.7)--(-2.85,2.4);
			\draw[<-] (-2.85,4.3)--(-2.85,3.6);
			\path [draw, ultra thick, blue] (-5,5)--(-3,5)--(-1,5);
			\path [draw, ultra thick, color=green!60!black] (-5,1)--(-3,1)--(-1,1);
			\path [draw, color=red] (-3,3)--(-1,3);
			\path [draw] (-3,1)--(-3,5);
			\draw[dashed] (-3,3) ellipse (1.3 and 2.3);
			\node at (-1,3)[below]{$k$};
			\node at (-5,5)[left]{$p_1$};
			\node at (-5,1)[left]{$p_2$};
			\node at (-1,5)[right]{$p_4$};
			\node at (-1,1)[right]{$p_3$};
			\node at (-2.8,4)[right]{$q_1$};
			\node at (-2.8,2)[right]{$q_2$};
		\end{tikzpicture}
	\end{gathered}
\end{equation}
with 
\begin{equation}\label{}
	p_1^2 = p_4^2 = -m_1^2\,,\qquad
	p_2^2 = p_3^2 = -m_2^2\,,\qquad
	k^2 = 0\,.
\end{equation}
The figure in \eqref{Aq1q2k} is meant to help remembering the definition of the ``transferred momenta'' $q_1$, $q_2$,
\begin{equation}\label{q1q2k}
	q_1 = p_1+p_4\,,\qquad
	q_2 = p_2+p_3\,,\qquad
	q_1+q_2+k=0
\end{equation}
and does not represent an actual topology.
Let us begin by discussing a useful choice of variables.

\subsection{Physical variables}
\label{ssec:physicalvariables}
We let
\begin{equation}\label{parq}
\begin{aligned}
	&p_1^\mu=-\bar m_1 u_1^\mu+q_1^\mu/2 \,,\qquad  p_4^\mu = \bar m_1 u_1^{\mu} + q_1^\mu/2 \\
	&p_2^\mu=-\bar m_2 u_2^{\mu} +q_2^\mu/2 \,,\qquad  p_3^\mu = \bar m_2 u_2^{\mu} + q_2^\mu/2 
\end{aligned}
\end{equation}
with 
\begin{equation}\label{}
	u_1^2 = -1 = u_2^2\,,\qquad
	y=-u_1\cdot u_2\ge1.
\end{equation}
In this way,
\begin{equation}\label{orth5}
	u_1\cdot q_1 = 0\,,\qquad u_2\cdot q_2=0\,,
\end{equation}
and
\begin{equation}\label{barmq1122}
	\bar m_1^2 = m_1^2 + \frac {q_1^2} 4, \qquad \bar m_2^2 = m_2^2 + \frac {q_2^2} 4\,.
\end{equation}
The momentum transfers $q_1^\mu$ and $q_2^\mu$ are not independent because of momentum conservation  \eqref{q1q2k} and of the mass-shell condition $k^2=0$, 
which implies
\begin{equation}\label{}
	q_1\cdot q_2 = -\frac12(q_1^2+q_2^2)\,.
\end{equation}
Five independent invariant Lorentz products can be taken as follows: 
\begin{equation}\label{barINV}
	y = -u_1\cdot u_2\,,\qquad
	\omega_{1}= u_{1} \cdot q_2\,,\qquad
	\omega_2=u_2\cdot q_1\,,\qquad
	q_{1}^2\,,\qquad
	q_2^2\,.
\end{equation}

The variable $y$ is the relative Lorentz factor of two observers with four-velocities $u_1^\mu$, $u_2^\mu$. Letting $\bar v$ denote the velocity of the former as seen from the rest frame of the latter (or vice-versa),
\begin{equation}\label{}
	y = \frac{1}{\sqrt{1-\bar v^2}}\,.
\end{equation}  
Using \eqref{q1q2k}, we see that $\omega_1$ and $\omega_2$ are the frequency of the graviton measured by these two observers,
\begin{equation}\label{}
	\omega_1= -u_1\cdot k\ge 0\,,\qquad
	\omega_2= -u_2\cdot k\ge 0\,.
\end{equation}

In order to simplify square roots that frequently appear in the calculations, it is convenient to define the following dimensionless variables
\begin{equation}\label{ratrel}
	x = y-\sqrt{y^2-1}\,,\qquad
	w_1 = \frac{\omega_1+\sqrt{\omega_1^2+q_2^2}}{q_2}\,,\qquad
	w_2 = \frac{\omega_2+\sqrt{\omega_2^2+q_1^2}}{q_1}\,,
\end{equation}
with the inverse relations given by
\begin{equation}\label{}
	y = \frac12\left(x+\frac{1}{x}\right)\!,\qquad
	\omega_1= \frac{q_2}{2}\left(w_1- \frac{1}{w_1}\right)\!,\qquad
	\omega_2= \frac{q_1}{2}\left(w_2-\frac{1}{w_2}\right)
\end{equation}
so that they obey the inequalities
\begin{equation}\label{}
	0<x\le 1\,,\qquad
	w_1\ge1\,,\qquad
	w_2\ge1\,.
\end{equation}
The limit $x\to 0$ corresponds to the high-energy (ultrarelativistic) regime and $x\to 1$ to the low-energy one. 

\subsection{Polarization tensor}
\label{ssec:polarization}

It is convenient to contract the amplitude with an appropriate polarization tensor, which we can build as follows.
We start from a vector 
\begin{equation}\label{polarizdecomp}
	\varepsilon^{\mu} = c_1 u_1^\mu + c_2 u_2^\mu + d_1 q_1^\mu + d_2 q_2^\mu
\end{equation}
with generic coefficients $c_1$, $c_2$, $d_1$, $d_2$.
We solve the transversality condition,
\begin{equation}\label{transversepsilon}
	k^\mu \varepsilon_\mu = - (q_1+q_2)^\mu \varepsilon_\mu = 0\,,
\end{equation}
by letting
\begin{equation}\label{physpol}
	d_1 = d_+ - d_-\,,\qquad d_2 = d_+ + d_-\,,\qquad
	d_- = \frac{c_1 \omega_1 + c_2 \omega_2}{q_1^2-q_2^2}\,.
\end{equation}

We then define the polarization tensor
\begin{equation}\label{ourpol}
	\varepsilon^{\mu\nu} = \varepsilon^\mu \varepsilon^\nu\,.
\end{equation}
This transverse, thanks to \eqref{physpol}. It can be also made traceless and related to more standard choices of graviton polarizations as detailed in Appendix~\ref{appendix:Kinematics}. 
We introduce the symbol
\begin{equation}\label{contrA1}
	\mathcal A_1^{} = \varepsilon_\mu \mathcal A_1^{\mu\nu} \varepsilon_\nu
\end{equation}
to denote the contracted amplitude.
Defining
\begin{equation}\label{hatvarepsilon}
	\hat\varepsilon^\mu   = \varepsilon^\mu - d_+ (q_1+q_2)^\mu=  \varepsilon^\mu  + d_+ k^\mu\,,
\end{equation}
gauge invariance requires that we can freely replace $\varepsilon^\mu$ with $\hat\varepsilon^\mu$
\begin{equation}\label{}
	\mathcal A_1 = \hat\varepsilon_\mu \mathcal A_1^{\mu\nu} \hat\varepsilon_\nu\,.
\end{equation}
The new polarization vector \eqref{hatvarepsilon} is independent of $d_+$, which thus constitutes a free parameter that ought to drop out from the final expression. This serves as a very useful cross-check of the calculations.

\section{Classical Limit and Integration}
\label{sec:classicallimit}

Let us spell out the decomposition of our amplitude in the \emph{classical} or \emph{near-forward} limit. 
In this regime the momentum transfers $q_1$, $q_2$ are taken to be simultaneously small with respect to the masses of the incoming particles, or, equivalently, the masses are taken to be large with respect to the exchanged momenta. We can therefore use a common scaling parameter as a bookkeeping device for the associated power counting. We shall use $\hbar$ for this bookkeeping purpose and define the scaling by
\begin{equation}\label{hbarscaling}
	q_{1,2}\sim \mathcal O(\hbar)\,,\qquad
	\bar m_{1,2} \sim u_{1,2}\sim \mathcal O(\hbar^0 )\,,\qquad 
	\varepsilon\sim\mathcal O(\hbar^0)\,,\qquad \hbar \to 0\,.
\end{equation}
We emphasize again that \eqref{hbarscaling} only serves to keep track of powers of the transferred momenta and does not refer to the actual dependence of the amplitude on the Planck constant after restoring standard units \cite{Kosower:2018adc}. 
We shall also scale the integrated momentum  $\ell$ associated to exchanged gravitons in the same way as the exchanged momenta $q_{1,2}$,
\begin{equation}\label{hbarscalingell}
	\ell \sim \mathcal O(\hbar)\,.
\end{equation} 
This enforces the expansion for the loop integrals in the soft region, which is the appropriate one to capture all the non-analytic dependence on $q_1$, $q_2$ in the amplitude.
From Eq.~\eqref{hbarscaling}, it also follows that
\begin{equation}\label{}
	k\sim \mathcal O(\hbar)\,,\qquad
	c_{1,2} \sim \mathcal O(\hbar^0)\,,\qquad
	d_{1,2} \sim \mathcal O(\hbar^{-1})\,,
\end{equation}
where $c_{1,2}$ and $d_{1,2}$ are the decomposition coefficients in \eqref{polarizdecomp}.

We follow the numbering of the 24 topologies, $\mathcal G_j$ with $j=1,\ldots,24$, associated to the integrand numerators given in Ref.~\cite{Carrasco:2021bmu}, which are depicted in Table~\ref{tab:diagrams}
\begin{table}[t]
	\begin{center}
		\begin{tabular}{| c | c | c | c |}
			\hline
			\begin{minipage}[c]{.2\textwidth}
				\vspace{10pt}
				\begin{center}
					\begin{tikzpicture}
						\path [draw, ultra thick, green!60!black] (-1,0)--(1,0);
						\path [draw, ultra thick, blue] (-1,1)--(1,1);
						\path [draw,thin] (.5,.97)--(.5,.03);
						\path [draw,thin] (-.5,.97)--(-.5,.03);
						\path [draw,thin] (0,.03)--(0,.23);
					\end{tikzpicture}
				\end{center}
				\vspace{3pt}
			\end{minipage}
			&
			\begin{minipage}[c]{.2\textwidth}
				\vspace{10pt}
				\begin{center}
					\begin{tikzpicture}
						\path [draw, ultra thick, green!60!black] (-1,0)--(1,0);
						\path [draw, ultra thick, blue] (-1,1)--(1,1);
						\path [draw,thin] (.5,.97)--(.5,.03);
						\path [draw,thin] (-.5,.97)--(-.5,.03);
						\path [draw,thin] (.5,.5)--(.7,.5);
					\end{tikzpicture}
				\end{center}
				\vspace{3pt}
			\end{minipage}
			&
			\begin{minipage}[c]{.2\textwidth}
				\vspace{10pt}
				\begin{center}
					\begin{tikzpicture}
						\path [draw, ultra thick, green!60!black] (-1,0)--(1,0);
						\path [draw, ultra thick, blue] (-1,1)--(1,1);
						\path [draw,thin] (.5,.97)--(.5,.03);
						\path [draw,thin] (-.5,.97)--(-.5,.03);
						\path [draw,thin] (.7,.03)--(.9,.23);
					\end{tikzpicture}
				\end{center}
				\vspace{3pt}
			\end{minipage}
			&
			\begin{minipage}[c]{.2\textwidth}
				\vspace{10pt}
				\begin{center}
					\begin{tikzpicture}
						\path [draw, ultra thick, green!60!black] (-1,.3)--(1,.3);
						\path [draw, ultra thick, blue] (-1,1)--(1,1);
						\path [draw,thin] (0,.97)--(0,.33);
						\draw[thin] (-.3,.27) arc (180:360:.3);
						\path [draw,thin] (0,-.03)--(.3,-.03);
					\end{tikzpicture}
				\end{center}
				\vspace{3pt}
			\end{minipage}
			\\
			$\mathcal{G}_1$, $P'$, $s_1=\tfrac12$
			&
			\cellcolor{white!30!yellow}
			$\mathcal{G}_2$,  $P$, $s_2=\tfrac12$
			&
			\cellcolor{white!30!yellow}
			$\mathcal{G}_3$, $P$, $s_3=1$
			&
			\cellcolor{yellow!80!red}
			$\mathcal{G}_4$,  $M'$, $s_4=\tfrac14$
			\\
			\hline
			\begin{minipage}[c]{.2\textwidth}
				\vspace{10pt}
				\begin{center}
					\begin{tikzpicture}
						\path [draw, ultra thick, green!60!black] (-1,0)--(1,0);
						\path [draw, ultra thick, blue] (-1,1)--(1,1);
						\path [draw,thin] (0,.97)--(0,.5);
						\path [draw,thin] (0,.5)--(-.5,.03);
						\path [draw,thin] (0,.5)--(.5,.03);
						\path [draw,thin] (0,.03)--(0,.23);
					\end{tikzpicture}
				\end{center}
				\vspace{3pt}
			\end{minipage}
			&
			\begin{minipage}[c]{.2\textwidth}
				\vspace{10pt}
				\begin{center}
						\begin{tikzpicture}
						\path [draw, ultra thick, green!60!black] (-1,.3)--(1,.3);
						\path [draw, ultra thick, blue] (-1,1)--(1,1);
						\path [draw,thin] (0,.97)--(0,.33);
						\draw[thin] (-.3,.27) arc (180:360:.3);
						\path [draw,thin] (.1,.33)--(.3,.43);
					\end{tikzpicture}
				\end{center}
				\vspace{3pt}
			\end{minipage}
			&
			\begin{minipage}[c]{.2\textwidth}
				\vspace{10pt}
				\begin{center}
					\begin{tikzpicture}
						\path [draw, ultra thick, green!60!black] (-1,0)--(1,0);
						\path [draw, ultra thick, blue] (-1,1)--(1,1);
						\path [draw,thin] (0,.97)--(0,.5);
						\path [draw,thin] (0,.5)--(-.5,.03);
						\path [draw,thin] (0,.5)--(.5,.03);
						\path [draw,thin] (-.5,1)--(-.75,.75);
					\end{tikzpicture}
				\end{center}
				\vspace{3pt}
			\end{minipage}
			&
			\begin{minipage}[c]{.2\textwidth}
				\vspace{10pt}
				\begin{center}
					\begin{tikzpicture}
						\path [draw, ultra thick, green!60!black] (-1,.3)--(1,.3);
						\path [draw, ultra thick, blue] (-1,1)--(1,1);
						\path [draw,thin] (0,.97)--(0,.33);
						\draw[thin] (-.3,.27) arc (180:360:.3);
						\path [draw,thin] (-.5,1)--(-.75,.75);
					\end{tikzpicture}
				\end{center}
				\vspace{3pt}
			\end{minipage}
			\\
			$\mathcal{G}_5$,  $P'$, $s_5=\tfrac14$
			&
			\cellcolor{white!60!red}
			$\mathcal{G}_6$,  $M$, $s_6=\tfrac12$
			&
			\cellcolor{white!30!yellow}
			$\mathcal{G}_7$,  $P$, $s_7=\tfrac12$
			&
			\cellcolor{white!60!red}
			$\mathcal{G}_8$, $M$, $s_8=\tfrac12$
			\\
			\hline
			\begin{minipage}[c]{.2\textwidth}
				\vspace{10pt}
				\begin{center}
					\begin{tikzpicture}
						\path [draw, ultra thick, green!60!black] (-1,0)--(1,0);
						\path [draw, ultra thick, blue] (-1,1)--(1,1);
						\path [draw,thin] (0,.97)--(0,.5);
						\path [draw,thin] (0,.5)--(-.5,.03);
						\path [draw,thin] (0,.5)--(.5,.03);
						\path [draw,thin] (.25,.25)--(.5,.5);
					\end{tikzpicture}
				\end{center}
				\vspace{3pt}
			\end{minipage}
			&
			\begin{minipage}[c]{.2\textwidth}
				\vspace{10pt}
				\begin{center}
					\begin{tikzpicture}
						\path [draw, ultra thick, green!60!black] (-1,0)--(1,0);
						\path [draw, ultra thick, blue] (-1,1)--(1,1);
						\path [draw,thin] (0,.97)--(0,.5);
						\path [draw,thin] (0,.5)--(-.5,.03);
						\path [draw,thin] (0,.5)--(.5,.03);
						\path [draw,thin] (.7,.03)--(.9,.23);
					\end{tikzpicture}
				\end{center}
				\vspace{3pt}
			\end{minipage}
			&
			\begin{minipage}[c]{.2\textwidth}
				\vspace{10pt}
				\begin{center}
				\begin{tikzpicture}
					\path [draw, ultra thick, green!60!black] (-1,.3)--(1,.3);
					\path [draw, ultra thick, blue] (-1,1)--(1,1);
					\path [draw,thin] (0,.97)--(0,.33);
					\draw[thin] (-.3,.27) arc (180:360:.3);
					\path [draw,thin] (.7,.33)--(.9,.53);
				\end{tikzpicture}
				\end{center}
				\vspace{3pt}
			\end{minipage}
			&
			\begin{minipage}[c]{.2\textwidth}
				\vspace{10pt}
				\begin{center}
					\begin{tikzpicture}
						\path [draw, ultra thick, green!60!black] (-1,.3)--(1,.3);
						\path [draw, ultra thick, blue] (-1,1)--(1,1);
						\path [draw,thin] (0,.97)--(0,.33);
						\draw[thin] (-.3,.27) arc (180:360:.3);
						\path [draw,thin] (0,.65)--(.3,.65);
					\end{tikzpicture}
				\end{center}
				\vspace{3pt}
			\end{minipage}
			\\
			\cellcolor{white!30!yellow}
			$\mathcal{G}_9$, $P$, $s_9=\tfrac12$
			&
			\cellcolor{white!30!yellow}
			$\mathcal{G}_{10}$,  $P$, $s_{10}=\tfrac12$
			&
			\cellcolor{white!60!red}
			$\mathcal{G}_{11}$, $M$, $s_{11}=\tfrac12$
			&
			\cellcolor{white!60!red}
			$\mathcal{G}_{12}$, $M$, $s_{12}=\tfrac12$
			\\
			\hline
			\begin{minipage}[c]{.2\textwidth}
				\vspace{10pt}
				\begin{center}
					\begin{tikzpicture}
						\path [draw, ultra thick, green!60!black] (-1,0)--(1,0);
						\path [draw, ultra thick, blue] (-1,1)--(1,1);
						\path [draw,thin] (0,.97)--(0,.5);
						\path [draw,thin] (0,.5)--(-.5,.03);
						\path [draw,thin] (0,.5)--(.5,.03);
						\path [draw,thin] (0,.75)--(.3,.75);
					\end{tikzpicture}
				\end{center}
				\vspace{3pt}
			\end{minipage}
			&
			\begin{minipage}[c]{.2\textwidth}
				\vspace{10pt}
				\begin{center}
					\begin{tikzpicture}
						\path [draw, ultra thick, green!60!black] (-1,.3)--(1,.3);
						\path [draw, ultra thick, blue] (-1,1)--(1,1);
						\path [draw,thin] (-.3,.97)--(-.3,.33);
						\draw[thin] (-.1,.27) arc (180:360:.3);
						\path [draw,thin] (.2,.33)--(.4,.43);
					\end{tikzpicture}
				\end{center}
				\vspace{3pt}
			\end{minipage}
			&
			\begin{minipage}[c]{.2\textwidth}
				\vspace{10pt}
				\begin{center}
					\begin{tikzpicture}
						\path [draw, ultra thick, green!60!black] (-1,0)--(1,0);
						\path [draw, ultra thick, blue] (-1,1)--(1,1);
						\path [draw,thin] (-.3,.97)--(-.3,.03);
						\draw[thin] (-.1,.03) arc (180:0:.3);
						\path [draw,thin] (.2,.33)--(.4,.43);
					\end{tikzpicture}
				\end{center}
				\vspace{3pt}
			\end{minipage}
			&
			\begin{minipage}[c]{.2\textwidth}
				\vspace{10pt}
				\begin{center}
					\begin{tikzpicture}
						\path [draw, ultra thick, green!60!black] (-1,0)--(1,0);
						\path [draw, ultra thick, blue] (-1,1)--(1,1);
						\path [draw,thin] (0,.03)--(0,.3);
						\path [draw,thin] (0,.97)--(0,.7);
						\draw[thin] (.2,.5) arc (0:360:.2);
						\draw[thin] (.2,.5)--(.5,.5);
					\end{tikzpicture}
				\end{center}
				\vspace{3pt}
			\end{minipage}
			\\
			\cellcolor{white!30!yellow}
			$\mathcal{G}_{13}$,  $P$, $s_{13}=\tfrac14$
			&
			\cellcolor{white!60!red}
			$\mathcal{G}_{14}$,  $M$, $s_{14}=\tfrac12$
			&
			\cellcolor{white!30!yellow}
			$\mathcal{G}_{15}$,  $P$, $s_{15}=\tfrac12$
			&
			\cellcolor{white!30!yellow}
			$\mathcal{G}_{16}$,  $P$, $s_{16}=\tfrac18$
			\\
			\hline
			\begin{minipage}[c]{.2\textwidth}
				\vspace{10pt}
				\begin{center}
					\begin{tikzpicture}
						\path [draw, ultra thick, green!60!black] (-1,0)--(1,0);
						\path [draw, ultra thick, blue] (-1,1)--(1,1);
						\path [draw,thin] (0,.03)--(0,.2);
						\path [draw,thin] (0,.97)--(0,.6);
						\draw[thin] (.2,.4) arc (0:360:.2);
						\path [draw,thin] (0,.8)--(.3,.8);
					\end{tikzpicture}
				\end{center}
				\vspace{3pt}
			\end{minipage}
			&
			\begin{minipage}[c]{.2\textwidth}
				\vspace{10pt}
				\begin{center}
					\begin{tikzpicture}
						\path [draw, ultra thick, green!60!black] (-1,0)--(1,0);
						\path [draw, ultra thick, blue] (-1,1)--(1,1);
						\path [draw,thin] (0,.03)--(0,.3);
						\path [draw,thin] (0,.97)--(0,.7);
						\draw[thin] (.2,.5) arc (0:360:.2);
						\path [draw,thin] (.5,.03)--(.7,.23);
					\end{tikzpicture}
				\end{center}
				\vspace{3pt}
			\end{minipage}
			&
			\begin{minipage}[c]{.2\textwidth}
				\vspace{10pt}
				\begin{center}
					\begin{tikzpicture}
					\path [draw, ultra thick, green!60!black] (-1,0)--(1,0);
					\path [draw, ultra thick, blue] (-1,1)--(1,1);
					\path [draw,thin] (-.5,.97)--(-.5,.03);
					\draw[thin] (-.3,.03) arc (180:0:.3);
					\path [draw,thin] (.5,.03)--(.7,.23);
				\end{tikzpicture}
				\end{center}
				\vspace{3pt}
			\end{minipage}
			&
			\begin{minipage}[c]{.2\textwidth}
				\vspace{10pt}
				\begin{center}
					\begin{tikzpicture}
						\node at (.03,1.04){$\cdots$};
						\path [draw,thin] (0,.03)--(0,.3);
						\path [draw,thin] (0,.97)--(0,.7);
						\draw[ultra thick] (.2,.5) arc (0:360:.2);
						\node at (.03,-.06){$\cdots$};
					\end{tikzpicture}
				\end{center}
				\vspace{3pt}
			\end{minipage}
			\\
			\cellcolor{white!30!yellow}
			$\mathcal{G}_{20}$,
			 $P$, $s_{20}=\tfrac14\times\tfrac12$
			&
			\cellcolor{white!30!yellow}
			$\mathcal{G}_{23}$, $P$, $s_{23}=\tfrac12\times \tfrac12$
			&
			\cellcolor{white!30!yellow}
			$\mathcal{G}_{24}$, $P$, $s_{24}=\tfrac12$
			&
			\cellcolor{white!70!black}
			$\mathcal G_{17,18,19,21,22}$
			\\
			\hline
		\end{tabular}
	\end{center}
	\caption{Topologies of the 24 numerators of Ref.~\cite{Carrasco:2021bmu}. Color code: yellow = pentagon ($P$), white = pentagon prime ($P'$), red = mushroom ($M$), orange = mushroom prime ($M'$), gray = quantum topologies. Our conventions on the external states are summarized in Eq.~\ref{Aq1q2k}.}
	\label{tab:diagrams}
\end{table}
and can be grouped into five families as in Table~\ref{tab:families}.
\begin{table}
	\begin{center}
		\begin{tabular}{| r | l |}
			\hline
			Family  & Topology \\
			\hline
			Pentagon ($P$) & 2, 3, 7, 9, 10, 13, 15, 16, 20, 23, 24\\
			Pentagon prime ($P'$) & 1, 5\\
			Mushroom ($M$) & 6, 8, 11, 12, 14\\
			Mushroom prime ($M'$) & 4\\
			\cdashline{1-2}
			Quantum & 17, 18, 19, 21, 22\\
			\hline
		\end{tabular}
		\caption{Families of topologies.}
		\label{tab:families}
	\end{center}
\end{table}
As we shall discuss in the Subsection~\ref{ssec:mapping}, all integrals belonging to the $P$, $P'$, $M$ and $M'$ families can be mapped to a collection of linearized pentagon integrals.
The 5 integrals of the quantum family are manifestly associated to intermediate processes, like creation of black-hole-/anti-black-hole pairs, which ought be disregarded in the classical limit, and indeed the associated integrals vanish in the soft region.
In this way, the 16 master integrals in Table~\ref{16MIt} below suffice to decompose the integrand via Integration By Parts and to evaluate the resulting integrals relevant for our purposes.

Each of the 16 numerators belonging the  $P$, $P'$, $M$ and $M'$ families should be multiplied by the appropriate propagators dictated by its diagram, and summed over the 8 independent permutations 
\begin{equation}\label{8permutations}
	P_8 = \left\{
	\sigma_1, \sigma_2, \sigma_3, \sigma_4, \sigma_2\sigma_3, \sigma_3\sigma_4, \sigma_2\sigma_4,\sigma_2\sigma_3\sigma_4
	\right\}
\end{equation}
generated by the following ones,
\begin{enumerate}
	\item[$\sigma_1$:] The trivial transformation (identity element).
	\item[$\sigma_2$:] The permutation interchanging the endpoints of the blue line in Eq.~\eqref{Aq1q2k}, sending $u_1^\mu \mapsto  -u_1^\mu$ and correspondingly\footnote{In principle also $x\mapsto-x$ corresponds to changing the sign of $y$, but the transformation in \eqref{perm2} is the one that leaves $\sqrt{y^2-1}=\tfrac12(\tfrac{1}{x}-x)$ invariant. For the same reason, one discards $w_1\mapsto-w_1$.}
	\begin{equation}\label{perm2}
		y \mapsto  -y\,,\qquad 
		\omega_1\to -\omega_1\,;\qquad
		x \mapsto -\frac1{x}\,,
		\qquad w_1 \mapsto \frac1{w_1}\,.
	\end{equation}
	\item[$\sigma_3$:] The permutation interchanging the endpoints of the green line in Eq.~\eqref{Aq1q2k}, sending $u_2^\mu \mapsto  -u_2^\mu$ and correspondingly
	\begin{equation}\label{perm3}
		y \mapsto  -y\,,\qquad 
		\omega_2\mapsto -\omega_2\,;
		\qquad
		x \mapsto -\frac1{x}\,,\qquad w_2 \mapsto \frac1{w_2}\,.
	\end{equation}
	\item[$\sigma_4$:] Particle-interchange symmetry, which corresponds to replacing the blue line with the green one and vicevesa, and to interchanging all particle labels $1\leftrightarrow2$.
\end{enumerate}
Of course, these operations should be performed while leaving $\varepsilon^\mu$ in Eq.~\eqref{polarizdecomp} invariant. For this reason, after expanding it as in \eqref{polarizdecomp}, one should compensate for the transformations of the basis vectors by also sending  $c_1\mapsto -c_1$ (resp. $c_2\mapsto -c_2$) when performing $\sigma_2$ (resp. $\sigma_3$) and by sending $c_1\leftrightarrow c_2$, $d_1 \leftrightarrow d_2$ when performing $\sigma_4$. Moreover, each diagram should only be summed over its nontrivial permutations. Equivalently, when summing over the whole $P_8$, each diagram should be supplied with the appropriate symmetry factor $s_i$ accounting for the fact that a subset of the permutations may leave it invariant. Massless bubble diagrams $\mathcal G_{20}$, $\mathcal G_{23}$ carry an additional factor of $1/2$ due to the freedom of relabeling the loop momentum.

After evaluating the integrals in each family using the integration measure
\begin{equation}\label{integrationmeasure}
	\int_\ell=
	e^{\gamma_E \epsilon} \int\frac{d^{4-2\epsilon}\ell}{i\pi^{2-\epsilon}}
\end{equation}
and summing over the allowed permutations, the last step is to multiply by the overall normalization factor $\mathcal N$ given by
\begin{equation}\label{overallfactor}
	\mathcal N = \frac{e^{-\gamma_E\epsilon}\mu^{2\epsilon}}{(4\pi)^{2-\epsilon}}(32\pi G)^{5/2} = \mathcal N_4 \,\bar \mu^{2\epsilon}\,,\qquad
	\mathcal N_4 = \frac{(32\pi G)^{5/2}}{(4\pi)^{2}}\,,\qquad
	\bar\mu^2= 4\pi e^{-\gamma_E} \mu^2 \,,
\end{equation}
with $\mu$ an arbitrary energy scale introduced by dimensional regularization.
All in all, we may summarize this construction as follows,
\begin{equation}\label{key}
	\mathcal A_1 = \mathcal N   \sum_{j=1}^{24}  \int_\ell \sum_{\sigma \in P_8} \sigma \left[s_j\,\frac{\mathcal G_j}{\text{den}_j} \right].
\end{equation}

\subsection{Mapping to the pentagon family}
\label{ssec:mapping}

In the limit \eqref{hbarscaling}, with an appropriate choice of loop momentum routing, all integrals in the pentagon family can be mapped to the following collection of integrals
\begin{equation}\label{jpentagon2}
	I_{i_1,i_2,i_3,i_4,i_5}
	=
	\int_\ell \frac{1}{(2 u_1\cdot \ell)^{i_1}(-2u_2\cdot \ell)^{i_2}(\ell^2)^{i_3} ((\ell+q_2)^2)^{i_4}((\ell-q_1)^2)^{i_5}}\,.
\end{equation}
In Eq.~\eqref{jpentagon2} and in the following, the $-i0$ prescription is left implicit for brevity.
The family \eqref{jpentagon2} is obtained from the conventional scalar pentagon with two massive lines (the momentum flows clockwise in the loop and, as in Eq.~\eqref{Aq1q2k}, the external momenta are all outgoing)
\begin{equation}\label{}
	\begin{gathered}
		\begin{tikzpicture}
			\path [draw, ultra thick, green!60!black] (-2,0)--(2,0);
			\path [draw, ultra thick, blue] (-2,3)--(2,3);
			\path [draw,thin] (-1.5,0)--(-1.5,3);
			\path [draw,thin] (1.5,0)--(1.5,3);
			\path [draw,thin] (1.5,1.5)--(2,1.5);
			\node at (2,1.5)[right]{$k$};
			\node at (-1.5,1.5)[left]{$\ell$};
			\node at (0,3)[below]{$-p_1+\ell$};
			\node at (0,0)[above]{$p_2+\ell$};
			\node at (1.5,2.5)[right]{$\ell-q_1$};
			\node at (1.5,0.5)[right]{$\ell+q_2$};
		\end{tikzpicture}
	\end{gathered}
\end{equation}
by linearizing the two massive propagators and factoring out $\bar{m}_1$, $\bar{m}_2$:
\begin{align}\label{}
	\frac{1}{(\ell-p_1)^2+m_1} &= \frac{1}{-2p_1\cdot \ell+\ell^2}= \frac{1}{2\bar m_1 u_1\cdot \ell +\ell^2-q_1^2} = \frac{1}{\bar m_1}\frac{1}{(2u_1\cdot \ell)} + \mathcal O(\hbar^0)\,,\\
		\frac{1}{(\ell+p_2)^2+m_2} &= \frac{1}{2p_2\cdot \ell+\ell^2}= \frac{1}{-2\bar m_2 u_2\cdot \ell +\ell^2+q_2^2} = \frac{1}{\bar m_2}\frac{1}{(-2u_2\cdot \ell)} + \mathcal O(\hbar^0)\,.
\end{align}
In our conventions, the sign of each propagator is fixed due to the $-i0$ prescription, e.g.
\begin{equation}\label{notequal}
 \frac{1}{(-2u_2\cdot \ell)} = \frac{1}{-2u_2\cdot \ell-i0}\,,\qquad  - \frac{1}{(2u_2\cdot \ell)}  = - \frac{1}{2u_2\cdot \ell-i0}\,,
\end{equation}
so that
\begin{equation}\label{mycut}
	\frac{1}{(-2u_2\cdot \ell)}+\frac1{(2u_2\cdot \ell)} = 2i\pi \delta(2u_2\cdot \ell)\,.
\end{equation}
A basis of for the family of integrals \eqref{jpentagon2}, which determine all the others via Integration By Parts (IBP), can be obtained using \texttt{LiteRed}~\cite{Lee:2012cn,Lee:2013mka} and is given by the 16 elements in Table \ref{16MIt} (although 7 of them can be deduced from the remaining 9 by using $\sigma_4$).
Using \texttt{HyperInt}~\cite{Panzer:2014caa} and dimensional shift identities \cite{Tarasov:1996br,Lee:2009dh,Lee:2010wea}, we have found the values of all such master integrals up to transcendental weight $2$. We present our results in Appendix~\ref{appendix:Masters} in the Euclidean region, and discuss their analytic continuation to the physical one in the Subsection~\ref{ssec:euclidean}.
\begin{table}[t]
	\begin{center}
		\begin{tabular}{| c | c | c | c |}
			\hline
			\begin{minipage}[c]{.2\textwidth}
				\vspace{10pt}
				\begin{center}
					\begin{tikzpicture}
						\path [draw, ultra thick, green!60!black] (-.65,0)--(.65,0);
						\path [draw, ultra thick, blue] (-.65,1)--(.65,1);
						\draw (0,.97) to[out=-60, in=60] (0,.03);
						\draw (0,.97) to[out=240, in=120] (0,.03);
						\draw (0,.03)--(.7,.5);
					\end{tikzpicture}
				\end{center}
				\vspace{3pt}
			\end{minipage}
			&
			\begin{minipage}[c]{.2\textwidth}
				\vspace{10pt}
				\begin{center}
					\begin{tikzpicture}
						\path [draw, ultra thick, green!60!black] (-.65,0)--(.65,0);
						\path [draw, ultra thick, blue] (-.65,1)--(.65,1);
						\draw (0,.97) to[out=-60, in=60] (0,.03);
						\draw (0,.97) to[out=240, in=120] (0,.03);
						\draw (0,.97)--(.7,.5);
					\end{tikzpicture}
				\end{center}
				\vspace{3pt}
			\end{minipage}
			&
			\begin{minipage}[c]{.2\textwidth}
				\vspace{10pt}
				\begin{center}
					\begin{tikzpicture}
						\path [draw, ultra thick, green!60!black] (-.65,0)--(-.5,.5)--(.65,0);
						\path [draw, ultra thick, blue] (-.65,1)--(-.5,.5)--(.65,1);
						\path [draw,thin] (.5,.05)--(.7,.5);
						\draw (-.5,.5) to[out=15, in=120] (.5,.05); 
					\end{tikzpicture}
				\end{center}
				\vspace{3pt}
			\end{minipage}
			&
			\begin{minipage}[c]{.2\textwidth}
				\vspace{10pt}
				\begin{center}
					\begin{tikzpicture}
						\path [draw, ultra thick, green!60!black] (-.65,0)--(-.5,.5)--(.65,0);
						\path [draw, ultra thick, blue] (-.65,1)--(-.5,.5)--(.65,1);
						\path [draw,thin] (.5,.95)--(.7,.5);
						\draw (-.5,.5) to[out=-15, in=240] (.5,.95); 
					\end{tikzpicture}
				\end{center}
				\vspace{3pt}
			\end{minipage}
			\\
			\cellcolor{green!70!yellow!40}
			$I_{0,0,1,0,1}$
			&
			\cellcolor{green!70!yellow!40}
			$I_{0,0,1,1,0}$
			&
			\cellcolor{green!50!gray!50}
			$I_{0,1,0,0,1}$
			&
			\cellcolor{green!50!gray!50}
			$I_{1,0,0,1,0}$
			\\
			\hline
			\begin{minipage}[c]{.2\textwidth}
				\vspace{10pt}
				\begin{center}
					\begin{tikzpicture}
						\path [draw, ultra thick, green!60!black] (-.65,0)--(.65,0);
						\path [draw, ultra thick, blue] (-.65,1)--(.65,1);
						\path [draw,thin] (0,.97)--(.5,.03);
						\path [draw,thin] (0,.97)--(-.5,.03);
						\path [draw,thin] (0,.97)--(.7,.5);
					\end{tikzpicture}
				\end{center}
				\vspace{3pt}
			\end{minipage}
			&
			\begin{minipage}[c]{.2\textwidth}
				\vspace{10pt}
				\begin{center}
					\begin{tikzpicture}
						\path [draw, ultra thick, green!60!black] (-.65,0)--(.65,0);
						\path [draw, ultra thick, blue] (-.65,1)--(.65,1);
						\path [draw,thin] (.5,.97)--(0,.03);
						\path [draw,thin] (-.5,.97)--(0,.03);
						\path [draw,thin] (0,.03)--(.7,.5);
					\end{tikzpicture}
				\end{center}
				\vspace{3pt}
			\end{minipage}
			&
			\begin{minipage}[c]{.2\textwidth}
				\vspace{10pt}
				\begin{center}
					\begin{tikzpicture}
						\path [draw, ultra thick, green!60!black] (-.65,0)--(.65,0);
						\path [draw, ultra thick, blue] (-.65,1)--(.65,1);
						\path [draw,thin] (0,.97)--(.5,.03);
						\path [draw,thin] (0,.97)--(-.5,.03);
						\path [draw,thin] (.5,.03)--(.7,.5);
					\end{tikzpicture}
				\end{center}
				\vspace{3pt}
			\end{minipage}
			&
			\begin{minipage}[c]{.2\textwidth}
				\vspace{10pt}
				\begin{center}
					\begin{tikzpicture}
						\path [draw, ultra thick, green!60!black] (-.65,0)--(.65,0);
						\path [draw, ultra thick, blue] (-.65,1)--(.65,1);
						\path [draw,thin] (.5,.97)--(0,.03);
						\path [draw,thin] (-.5,.97)--(0,.03);
						\path [draw,thin] (.5,.97)--(.7,.5);
					\end{tikzpicture}
				\end{center}
				\vspace{3pt}
			\end{minipage}
			\\
			\cellcolor{green!70!yellow!40}
			$I_{0,1,1,1,0}$
			&
			\cellcolor{green!70!yellow!40}
			$I_{1,0,1,0,1}$
			&
			\cellcolor{green!70!yellow!40}
			$I_{0,1,1,0,1}$
			&
			\cellcolor{green!70!yellow!40}
			$I_{1,0,1,1,0}$
			\\
			\hline
			\begin{minipage}[c]{.2\textwidth}
				\vspace{10pt}
				\begin{center}
					\begin{tikzpicture}
						\path [draw, ultra thick, green!60!black] (-.65,0)--(-.5,.5)--(.65,0);
						\path [draw, ultra thick, blue] (-.65,1)--(-.5,.5)--(.65,1);
						\path [draw,thin] (.5,.95)--(.5,.037);
						\path [draw,thin] (.5,.95)--(.7,.5);
					\end{tikzpicture}
				\end{center}
				\vspace{3pt}
			\end{minipage}
			&
			\begin{minipage}[c]{.2\textwidth}
				\vspace{10pt}
				\begin{center}
					\begin{tikzpicture}
						\path [draw, ultra thick, green!60!black] (-.65,0)--(-.5,.5)--(.65,0);
						\path [draw, ultra thick, blue] (-.65,1)--(-.5,.5)--(.65,1);
						\path [draw,thin] (.5,.95)--(.5,.037);
						\path [draw,thin] (.5,.037)--(.7,.5);
					\end{tikzpicture}
				\end{center}
				\vspace{3pt}
			\end{minipage}
			&
			\begin{minipage}[c]{.2\textwidth}
				\vspace{10pt}
				\begin{center}
					\begin{tikzpicture}
						\path [draw, ultra thick, green!60!black] (-.65,0)--(.65,0);
						\path [draw, ultra thick, blue] (-.65,1)--(.65,1);
						\path [draw,thin] (0,.97)--(.5,.03);
						\path [draw,thin] (0,.97)--(-.5,.03);
						\path [draw,thin] (.25,.5)--(.7,.5);
					\end{tikzpicture}
				\end{center}
				\vspace{3pt}
			\end{minipage}
			&
			\begin{minipage}[c]{.2\textwidth}
				\vspace{10pt}
				\begin{center}
					\begin{tikzpicture}
						\path [draw, ultra thick, green!60!black] (-.65,0)--(.65,0);
						\path [draw, ultra thick, blue] (-.65,1)--(.65,1);
						\path [draw,thin] (.5,.97)--(0,.03);
						\path [draw,thin] (-.5,.97)--(0,.03);
						\path [draw,thin] (.25,.5)--(.7,.5);
					\end{tikzpicture}
				\end{center}
				\vspace{3pt}
			\end{minipage}
			\\
			\cellcolor{green!50!gray!50}
			$I_{1,1,0,1,0}$
			&
			\cellcolor{green!50!gray!50}
			$I_{1,1,0,0,1}$
			&
			\cellcolor{green!70!yellow!40}
			$I_{0,1,1,1,1}$
			&
			\cellcolor{green!70!yellow!40}
			$I_{1,0,1,1,1}$
			\\
			\hline
			\begin{minipage}[c]{.2\textwidth}
				\vspace{10pt}
				\begin{center}
					\begin{tikzpicture}
						\path [draw, ultra thick, green!60!black] (-.65,0)--(-.5,.5)--(.65,0);
						\path [draw, ultra thick, blue] (-.65,1)--(-.5,.5)--(.65,1);
						\path [draw,thin] (.5,.95)--(.5,.035);
						\path [draw,thin] (.5,.5)--(.7,.5);
					\end{tikzpicture}
				\end{center}
				\vspace{3pt}
			\end{minipage}
			&
			\begin{minipage}[c]{.2\textwidth}
				\vspace{10pt}
				\begin{center}
					\begin{tikzpicture}
						\path [draw, ultra thick, green!60!black] (-.65,0)--(.65,0);
						\path [draw, ultra thick, blue] (-.65,1)--(.65,1);
						\path [draw,thin] (.5,.97)--(.5,.03);
						\path [draw,thin] (-.5,.97)--(-.5,.03);
						\path [draw,thin] (.5,.03)--(.7,.5);
					\end{tikzpicture}
				\end{center}
				\vspace{3pt}
			\end{minipage}
			&
			\begin{minipage}[c]{.2\textwidth}
				\vspace{10pt}
				\begin{center}
					\begin{tikzpicture}
						\path [draw, ultra thick, green!60!black] (-.65,0)--(.65,0);
						\path [draw, ultra thick, blue] (-.65,1)--(.65,1);
						\path [draw,thin] (.5,.97)--(.5,.03);
						\path [draw,thin] (-.5,.97)--(-.5,.03);
						\path [draw,thin] (.5,.97)--(.7,.5);
					\end{tikzpicture}
				\end{center}
				\vspace{3pt}
			\end{minipage}
			&
			\begin{minipage}[c]{.2\textwidth}
				\vspace{10pt}
				\begin{center}
					\begin{tikzpicture}
						\path [draw, ultra thick, green!60!black] (-.65,0)--(.65,0);
						\path [draw, ultra thick, blue] (-.65,1)--(.65,1);
						\path [draw,thin] (.5,.97)--(.5,.03);
						\path [draw,thin] (-.5,.97)--(-.5,.03);
						\path [draw,thin] (.5,.5)--(.7,.5);
					\end{tikzpicture}
				\end{center}
				\vspace{3pt}
			\end{minipage}
			\\
			\cellcolor{green!50!gray!50}
			$I_{1,1,0,1,1}$
			&
			\cellcolor{green!70!yellow!40}
			$I_{1,1,1,0,1}$
			&
			\cellcolor{green!70!yellow!40}
			$I_{1,1,1,1,0}$
			&
			\cellcolor{green!70!yellow!40}
			$I_{1,1,1,1,1}$
			\\
			\hline
		\end{tabular}
	\end{center}
	\caption{Topologies of the 16 master integrals for the pentagon family. Color code: lighter green = non-analytic in $q^2$, darker green = analytic in $q^2$. The appearance of the latter type of topologies where matter lines ``touch'' and which do not appear in Table~\ref{tab:diagrams} is induced by the IBP reduction, whose coefficients can be non-analytic in $q^2$ and  thus induce long-range effects in position space. Such contributions would be scale-less in the $2\to2$ kinematics.}
	\label{16MIt}
\end{table}

It turns out that the pentagon prime, mushroom and mushroom prime families can be also mapped to the integrals \eqref{jpentagon2} in the limit \eqref{hbarscaling}, by suitably decomposing the linearized propagators into partial fractions and applying symmetry transformations.
Let us discuss this step in detail focusing on a prototypical integral for each family, with propagators raised to the first power. Generalizing this procedure to any other positive power is straightforward.
A typical integral of the pentagon prime family takes the form
\begin{equation}\label{}
	\begin{gathered}
		\begin{tikzpicture}
			\path [draw, ultra thick, green!60!black] (-.65,0)--(.65,0);
			\path [draw, ultra thick, blue] (-.65,1)--(.65,1);
			\path [draw,thin] (.5,.97)--(.5,.03);
			\path [draw,thin] (-.5,.97)--(-.5,.03);
			\path [draw,thin] (0,.03)--(0,.3);
			\node at (-.5,.5)[left]{$\ell$};
		\end{tikzpicture}
	\end{gathered}\to
	I^{(P')} 
=\int_\ell \frac{1}{(2u_1\cdot \ell)(-2u_2\cdot \ell)(-2u_2\cdot \ell+2\omega_2)\ell^2(\ell-q_1)^2}
\end{equation}
and decomposing the second and third propagator into partial fractions leads to
\begin{equation}\label{}
	I^{(P')} = \frac{1}{2\omega_2}\int_\ell \frac{1}{(2u_1\cdot \ell)(-2u_2\cdot \ell)\ell^2(\ell-q_1^2)} - \frac{1}{2\omega_2}\int_\ell \frac{1}{(2u_1\cdot \ell)(-2u_2\cdot \ell+2\omega_2)\ell^2(\ell-q_1)^2}\,.
\end{equation}
We can change integration variable in the second integral, letting $\ell \to q_1-\ell$, and obtain
\begin{equation}\label{key}
	I^{(P')} = \frac{1}{2\omega_2}\int_\ell \frac{1}{(2u_1\cdot \ell)(-2u_2\cdot \ell)\ell^2(\ell-q_1^2)} - \frac{1}{2\omega_2}\int_\ell \frac{1}{(-2u_1\cdot \ell)(2u_2\cdot \ell)\ell^2(\ell-q_1)^2}\,.
\end{equation}
The two integrals do not simply cancel against each other, due to \eqref{notequal}, \eqref{mycut}, but we can map them to the family \eqref{jpentagon2} by applying permutations $\sigma_2$ and $\sigma_3$ to the second one,
\begin{equation}\label{P'PF}
	I^{(P')} = \frac{1}{2\omega_2}\, I_{1,1,1,0,1}- \frac{1}{2\omega_2} \,\sigma_2\sigma_3 I_{1,1,1,0,1}\,.
\end{equation}
For a typical integral of the mushroom family,\footnote{For the diagram in the figure $1/(2u_2\cdot \ell)$ would actually be raised to the second power.}
\begin{equation}\label{}
		\begin{gathered}
		\begin{tikzpicture}
			\path [draw, ultra thick, green!60!black] (-.65,0)--(0,.98)--(.65,0);
			\path [draw, ultra thick, blue] (-.65,1)--(.65,1);
			\path [draw,thin] (.5,.17)--(-.5,.17);
			\path [draw,thin] (.3,.58)--(.55,.58);
			\node at (0,0){$\ell$};
		\end{tikzpicture}
	\end{gathered}
\to
	I^{(M)} = \int_\ell \frac{1}{(2u_2\cdot \ell)(2u_2\cdot \ell-2\omega_2)\ell^2}
\end{equation}
we have
\begin{equation}\label{}
	I^{(M)} = - \frac{1}{2\omega_2} \int_\ell \frac{1}{(2u_2\cdot \ell)\ell^2} + \frac{1}{2\omega_2} \int_\ell \frac{1}{(2u_2\cdot \ell-2\omega_2)\ell^2}\,.
\end{equation}
Noting that the first integral on the right-hand side is scaleless, and sending $\ell\to q_1-\ell$ in the second one, we find
\begin{equation}\label{}
	I^{(M)} = \frac{1}{2\omega_2}\,I_{0,1,0,0,1}\,.
\end{equation}
Finally, a typical mushroom prime integral takes the form
\begin{equation}\label{}
	\begin{gathered}
		\begin{tikzpicture}
			\path [draw, ultra thick, green!60!black] (-.65,0)--(0,.98)--(.65,0);
			\path [draw, ultra thick, blue] (-.65,1)--(.65,1);
			\path [draw,thin] (.5,.17)--(-.5,.17);
			\path [draw,thin] (0,-.1)--(0,.17);
			\node at (-.4,.6)[left]{$\ell+p_3$};
		\end{tikzpicture}
	\end{gathered}
\to
	I^{(M')} = \int_\ell \frac{1}{(2u_2\cdot \ell)(2u_2\cdot \ell-2\omega_2)(\ell-q_1)^2(\ell+q_2)^2}\,,
\end{equation}
and leads to the following partial fractions
\begin{equation}\label{}
	I^{(M')} = - \frac{1}{2\omega_2} \int_\ell \frac{1}{(2u_2\cdot \ell)(\ell-q_1)^2(\ell+q_2)^2} +  \frac{1}{2\omega_2}  \int_\ell \frac{1}{(2u_2\cdot \ell-2\omega_2)(\ell-q_1)^2(\ell+q_2)^2}\,.
\end{equation}
Performing the permutation $\sigma_3$ in the first integral and sending $\ell\to q_1-q_2-\ell$ in the second one,
\begin{equation}\label{M'MF}
	I^{(M')} = -\frac{1}{2\omega_2}\,   \sigma_3I_{0,1,0,1,1} +  \frac{1}{2\omega_2} \,I_{0,1,0,1,1}\,.
\end{equation} 
Let us comment that, since one is ultimately summing over all allowed permutations \eqref{jpentagon2} to build the full integrand from the 19 diagrams in Table~\ref{tab:diagrams}, it is not strictly necessary to apply $\sigma_2\sigma_3$ as in \eqref{P'PF} and $\sigma_3$ as in \eqref{M'MF}. One can also first treat the two contributions to each equation as separate objects, perform the IBP reduction and mapping to the master integrals only for the appropriate permutation, and then perform all 8 permutations on the result.

\subsection{Euclidean variables and analytic continuation}
\label{ssec:euclidean}

The amplitude is the boundary value of an analytic function which develops branch cuts when its variables are located in the physical region. For this reason it is convenient to introduce complex variables $y_E$, $\omega_{1E}$, $\omega_{2E}$ such that the physical region corresponds to setting
\begin{equation}\label{yEom1Eom2E}
	y_E =- y-i0\,,\qquad \omega_{1E}=-\omega_1-i0\,,\qquad \omega_{2E}=-\omega_2-i0\,,
\end{equation}
with $y$, $\omega_1$, $\omega_2$ the invariant products defined in Eq.~\eqref{barINV}.
The new variables allow for manifestly real expressions of the master integrals \eqref{jpentagon2} when they are taken in the Euclidean region, defined by
\begin{equation}\label{Euclideanregion}
	y_E\ge1\,,\qquad \omega_{1E}\ge 0\,,\qquad \omega_{2E}\ge0\,.
\end{equation}
We similarly define new rationalized variables according to
\begin{equation}\label{}
	x_E = y_E+ \sqrt{y_E^2-1}\,,\qquad
	w_{1E} = \frac{\sqrt{\omega_{1E}^2+q_2^2}+\omega_{1E}}{q_2}\,,\qquad
	w_2 = \frac{\sqrt{\omega_{2E}^2+q_1^2}+\omega_{2E}}{q_1}\,,
\end{equation}
so that
\begin{equation}\label{}
	y_E = \frac12\left(x_E+\frac{1}{x_E}\right)\!,\qquad
	\omega_{1E}= \frac{q_2}{2}\left(w_{1E}-\frac{1}{w_{1E}}\right)\!,\qquad
	\omega_{2E}= \frac{q_1}{2}\left(w_{2E}-\frac{1}{w_{2E}}\right).
\end{equation}
The rationalized variables fall in the physical region when
\begin{equation}\label{xEx}
	x_E = -x+i0\,,\qquad
	w_{1E} = -w_1-i0\,,\qquad
	w_{2E} = -w_2-i0\,,
\end{equation}
with $x$, $w_1$, $w_2$ as in Eq.~\eqref{ratrel}.
The conditions \eqref{Euclideanregion} defining the Euclidean region instead translate to the following ones in terms of the rationalized variables,
\begin{equation}\label{Euclideanregionrat}
	x_E\ge1\,,\qquad
	w_{1E}\ge1\,,\qquad
	w_{2E}\ge 1\,.
\end{equation}

Mapping back to the physical variables discussed in Subsection~\ref{ssec:physicalvariables} via \eqref{yEom1Eom2E}, \eqref{xEx}, one encounters branch singularities when the variables fall in the physical region. We have expressed our  master integrals (see Appendix~\ref{appendix:Masters}) in terms of the  analytic functions
\begin{equation}\label{}
\begin{split}
	&\log x_E\,,\quad \log(x_E-1)\,,\quad \log w_{1E}\,,\quad \log(w_{1E}\pm 1)\,,\quad \log w_{2E}\,,\quad \log(w_{2E}\pm 1)\,,
\end{split}
\end{equation}
and
\begin{equation}\label{}
	\operatorname{Li}_2\left(\pm\frac{1}{x_E}\right),\qquad \operatorname{Li}_2\left(\pm\frac{1}{w_{1E}}\right),\qquad
	\operatorname{Li}_2\left(\pm\frac{1}{w_{2E}}\right)
\end{equation}
so that their expressions are manifestly real in the Euclidea region \eqref{Euclideanregionrat}.
We then perform the analytic continuation back to the physical region \eqref{xEx} by letting
\begin{equation}\label{analyticcontinuationLOGs}
	\begin{split}
\log x_E  &\to \log x + i q_I \pi\,,\\
\log w_{1E} &\to \log w_{1}  - i q_O \pi\,,\\
\log w_{2E} &\to \log w_{2}  - i q_A \pi\,,
	\end{split}
\end{equation}
where we have kept all analytic continuations in principle arbitrary.
We find that, consistently with the $i0$ prescriptions in \eqref{xEx} or equivalently by demanding consistency with the exponentiation of infrared divergences (see Subsection~\ref{ssec:IR}), 
\begin{equation}\label{qSqC}
	q_I = q_O = q_A = +1\,.
\end{equation}
The sign of the $i0$ prescription $q_I$ for $x_E$ matches the elastic calculation in \cite{DiVecchia:2021bdo} where for instance in the double box solution $\log x_E \to \log x +i\pi$.
The elementary choices made in \eqref{analyticcontinuationLOGs} then resolve all other branch ambiguities upon reducing the analytic continuation of the dilogarithms to those of conventional logarithms via
\begin{equation}\label{}
	\operatorname{Li}_2\left(\frac{1}{z}\right) = - \operatorname{Li}_2(z) - \frac{\pi^2}{6} - \frac{1}{2}(\log (-z))^2\,,
\end{equation}
which holds whenever $z$ doesn't belong to the positive real axis.

\section{Structure of the Amplitude in the Classical Limit}
\label{sec:structure}

Looking at the integrand obtained by combining the diagrams in Table~\ref{tab:diagrams}, we find the following structure in the limit \eqref{hbarscaling}, \eqref{hbarscalingell} (including the scaling of the measure element $d^4\ell$),
\begin{equation}\label{}
	\mathcal A_1^{} = \mathcal A_1^{[-3]} + \mathcal A_1^{[-2]} + \mathcal A_1^{[-1]} + \mathcal O(\hbar^0)\,,
\end{equation}
where
\begin{equation}\label{}
	\mathcal {A}^{[-j]} \sim \mathcal O(\hbar^{-j})\,.
\end{equation}
We expand each coefficient for small $\epsilon=\frac{4-D}{2}$, defining
\begin{equation}\label{}
	\mathcal A_1^{[-j]} = \bar\mu^{2\epsilon}\left[\frac{\mathcal A_1^{[-j,-2]}}{\epsilon^2} + \frac{\mathcal A_1^{[-j,-1]}}{\epsilon}  + \mathcal A_1^{[-j,0]} + \mathcal O(\epsilon)\right].
\end{equation}
The first (second) index within square brackets thus refers to the  $\hbar$ (resp.~$\epsilon$) scaling.

After inputting the values of the master integrals, we find that (for nonzero graviton frequencies)
\begin{equation}\label{hbm3}
	\mathcal A_1^{[-3]} = \mathcal O(\epsilon)\,.
\end{equation}
This cancellation is expected because it mirrors a similar one occurring for the tree-level amplitude $\mathcal A_0$, whose classical limit also naively goes like $\hbar^{-3}$ (and it indeed involves terms localized at zero frequency at that order), while its actual scaling is $\hbar^{-2}$ in our present conventions. Eq.~\eqref{hbm3} also serves a nontrivial check of the symmetry factors because it relies on $s_3 = 2s_1$ and on $s_6+s_{24}=s_{11}+s_{14}$. 
We also find that the coefficient of the double pole in $\epsilon$ vanishes,
\begin{equation}\label{am2m2}
	\mathcal A_1^{[-j,-2]}=0\,,\qquad j=2,1\,.
\end{equation}
For instance, both $\mathcal G_{2}$, $\mathcal G_4$ and $\mathcal G_{9}$ would naively diverge like $1/\epsilon^2$ to order $\mathcal O(\hbar^{-2})$, but thanks to the transversality condition \eqref{physpol} such divergences cancel between $\mathcal G_2$ and $\mathcal G_9$, and separately in $\mathcal G_4$. This serves as a cross check that $s_2=s_9$. This cancellation is also expected on general grounds \cite{Weinberg:1965nx} as we shall discuss more in detail shortly.

Taking into account the vanishing of these coefficients, we find the following structure of the amplitude in the classical limit,  
\begin{equation}\label{structureA1}
\mathcal A_{1} = \bar{\mu}^{2\epsilon} \left\{\left[\frac{\mathcal A_1^{[-2,-1]}}{\epsilon} + \mathcal A_1^{[-2,0]}  \right] + \left[\frac{\mathcal A_1^{[-1,-1]}}{\epsilon} + \mathcal A_1^{[-1,0]} \right]
+ \mathcal O(\epsilon)+\mathcal O(\hbar^0) \right\}.
\end{equation}
The functions $\mathcal A_1^{[-2,-1]}$, $\mathcal A_1^{[-2,0]}$, $\mathcal A_1^{[-1,-1]}$, $\mathcal A_1^{[-1,0]}$ constitute the main results of the present work and are all provided in the ancillary files in attachment, where for convenience we collect their expressions after dividing by $\mathcal N_4$ defined in \eqref{overallfactor}.
We have checked that our expressions for the classical terms, $\mathcal A_1^{[-1,-1]}$ and $\mathcal A_1^{[-1,0]}$, agree with the results of Ref.~\cite{Radu&Co} on numerical points, up to an  overall sign.
In its turn, this also ensures agreement wit the results of Ref.~\cite{QMUL}.
The remainder of this section is devoted to the discussion and illustration of Eq.~\eqref{structureA1}.

We find that the coefficients of the $\epsilon^{-1}$ poles, $\mathcal A_1^{[-2,-1]}$, $\mathcal A_1^{[-1,-1]}$, are in complete agreement with the prediction obtained from the exponentiation of infrared divergences, which fixes them completely in terms of the tree-level  five-point amplitude $\mathcal A_{0}$ times a universal factor \cite{Weinberg:1965nx}. The combinations $\mathcal A_1^{[-2,-1]}/\mathcal N_4$ and $\mathcal A_1^{[-1,-1]}/\mathcal N_4$ have uniform transcendental weight $1$, i.e.~they are rational functions of the invariants $x$, $w_1$, $w_2$ in \eqref{ratrel}, $q_1$ and $q_2$ times $i\pi$.
For all terms displayed in \eqref{structureA1}, we also find agreement with Weinberg's soft theorem \cite{Weinberg:1964ew}, which dictates that, as the frequency of the graviton tends to zero, their most singular term must reduce to a universal factor times the one-loop four-point amplitude $\mathcal A_1^{(4)}$.
Moreover, the ``superclassical'' terms $\mathcal A_1^{[-2,-1]}$, $\mathcal A_1^{[-2,0]}$, as well as $\mathcal A_1^{[-1,-1]}$ and the terms proportional to the imaginary unit in $\mathcal A_1^{[-1,0]}$, when written in terms of $c_1$ and $c_2$, arise from the cuts of the amplitude exactly as predicted by unitarity.

The terms that do not multiply the imaginary unit in $\mathcal A_1^{[-1,0]}/\mathcal N_4$ have uniform transcendental weight $2$. When expressed in terms of \begin{equation}\label{thelogs}
 \log x\,,\qquad \log(1\pm x)\,,\qquad \log w_{1,2}\,,\qquad \log(w_{1,2}\pm 1)\,,\qquad \log(q_{1,2})
\end{equation} 
and 
\begin{equation}\label{thedilogs}
	\operatorname{Li}_2(x)\,,\qquad \operatorname{Li}_2\left(\frac{1}{w_{1,2}}\right)\,,
\end{equation}
all dependence on the logarithms and dilogarithms drops out in such terms, so that they reduce to rational functions of the invariants $x$, $w_1$, $w_2$ in \eqref{ratrel}, $q_1$ and $q_2$ times $\pi^2$. After this simplification, the structure of this piece is thus analogous to that of the $\mathcal O(\hbar^{-1})$ in the elastic $2\to2$ amplitude at one loop (Eq.~\eqref{triangles} below).

The terms that multiply the imaginary unit in $\mathcal A_1^{[-1,0]}/\mathcal N_4$ (and similarly the combination  $\mathcal A_1^{[-2,0]}/\mathcal N_4$), instead, can have transcendental weight either $2$, i.e.~reduce to rational functions times the logarithms \eqref{thelogs} times $i\pi$, or $1$, i.e.~reduce to rational functions times $i\pi$.
Schematically,
\begin{equation}\label{}
	\frac{\mathcal A_1^{[-1,0]}}{\mathcal N_4} = \pi^2 
	Q(x,w_1,w_2,q_1,q_2)
	+
	i\pi \Big[
	\sum_j (\log x_j)\, R_j(x,w_1,w_2,q_1,q_2)
	+
	S(x,w_1,w_2,q_1,q_2)
	\Big]
\end{equation}
where $Q$, $R_j$ and $S$ are real, rational functions of the invariants and $\log x_j$ are the logarithms \eqref{thelogs}.

\subsection{Exponentiation of infrared divergences}
\label{ssec:IR}

Infrared divergences in gravity amplitudes follow a simple exponential patter first clarified by Weinberg \cite{Weinberg:1965nx} (see also \cite{Ware:2013zja,Heissenberg:2021tzo}), 
\begin{equation}\label{}
	\mathcal A_{\alpha\to\beta} = e^{\mathcal W_{\alpha\to\beta}} [\mathcal A_{\alpha\to\beta}]_\text{\tiny IR finite}
\end{equation}
with $\mathcal W_{\alpha\to\beta}$ an infrared-divergent, one-loop-exact exponent whose expression in terms of the states $\alpha$, $\beta$ takes a universal form.
Accordingly, the infrared divergences of any one-loop five-point amplitude are equal to
\begin{equation}\label{}
	\mathcal W = \frac{G}{2\pi \epsilon} \sum_{n,m=1}^5 w_{nm}
\end{equation}
times the tree-level five-point amplitude with the same external states.
When the 5 particles are massive, letting ($\eta_n$ is +1 if $n$ is outgoing and $-1$ if $n$ is incoming)
\begin{equation}\label{calW}
	\zeta_{nm} = - \eta_n \eta_m p_n\cdot p_m > 0  \,,
\end{equation}
we have
\begin{equation}\label{wnm}
	w_{nm} = \frac{\zeta_{nm}^2-\tfrac 12 \,m_n^2m_m^2}{\sqrt{\zeta_{nm}^2-m_n^2m_m^2}}
	\left[
	\eta_n\eta_m 
	\log\frac{\zeta_{nm}+\sqrt{\zeta_{nm}^2-m_n^2m_m^2}}{\zeta_{nm}-\sqrt{\zeta_{nm}^2-m_n^2m_m^2}}
	- i\pi \eta_{nm}
	\right]
\end{equation}
where $\eta_{nm} = +1$ provided $n\neq m$ and $n$ and $m$ are both outgoing or both incoming, and vanishes otherwise.
Moreover $w_{nn} = m_n^2/2$.
When $m_5\to0$ and $p_5\to k$, the function $\mathcal W$ is smooth and reduces to
\begin{equation}\label{calW5}
	\mathcal W = \frac{G}{2\pi \epsilon} \sum_{n,m=1}^4 w_{nm}  + \frac{G}{2\pi \epsilon} \sum_{n=1}^42(-p_n\cdot k)\left[\log\frac{4(p_n\cdot k)^2}{\Lambda^2 m_n^2}-i\pi \eta_{n5}\right]
\end{equation}
with $\Lambda$ an arbitrary energy scale.
One can explicitly check this by taking the limit in \eqref{calW} and by using momentum conservation
\begin{equation}\label{}
	p_1+p_2+p_4+p_4 = -k
\end{equation}
to show that a potentially dangerous $\log m_5$ cancels out, leaving behind an arbitrary reference scale $\Lambda$ in the logarithm in \eqref{calW5}.
All in all, this dictates the IR divergences of our amplitude,
\begin{equation}\label{W65}
	\mathcal A_1 = \mathcal W \mathcal A_0 + \mathcal O(\epsilon^0)\,,
\end{equation}
with $\mathcal A_0 = \varepsilon_\mu \mathcal A_0^{\mu\nu} \varepsilon_\nu$ the tree-level five-point amplitude \eqref{A05pt}. In particular, \eqref{W65} predicts that no double pole $1/\epsilon^2$ should occur and Eq.~\eqref{am2m2} is consistent with this prediction.

We checked that, expanding to leading and subleading
order in $\hbar$, 
\begin{equation}\label{}
	\mathcal W = \mathcal W^{[0]} +
	\mathcal W^{[1]} + \mathcal O(\hbar^2)\,,
\end{equation}
where
\begin{equation}\label{W0W1}
	\mathcal W^{[0]}
	=
	-\frac{i  2G \bar m_1 \bar m_2 \left(y ^2-\tfrac12\right)}{\epsilon\, \sqrt{y ^2-1}}\,,
	\qquad
	\mathcal W^{[1]} = - \frac{i G}{\epsilon}  (\bar m_1 \omega_1+ \bar m_2 \omega_2)\,.
\end{equation}
Therefore, to this order in $\hbar$, the Weinberg exponent is purely imaginary. $\mathcal W^{[0]}$ is a divergent phase arising from soft graviton exchanges between lines $1-2$ and $3-4$ while $\mathcal W^{[1]}$ arises from those between the outgoing graviton line and lines $3$, $4$. 
The other ingredient is the tree-level five-point amplitude, which in the classical limit is given by 
$\mathcal A_0 = \mathcal A_0^{[-2]} + \mathcal O(\hbar^{0})$ as in \eqref{CLtree} \cite{Goldberger:2016iau,Luna:2017dtq,Jakobsen:2021lvp,Mougiakakos:2021ckm,Manohar:2022dea}.
Note the lack of $\mathcal O(\hbar^{-1})$ corrections to this leading $\mathcal O(\hbar^{-2})$ result, as discussed below Eq~\eqref{CLtree}.
Consistently with these general facts, we checked that our result \eqref{structureA1} obeys
\begin{equation}\label{am2m1W65}
	\frac1\epsilon\,
	\mathcal A_1^{[-2,-1]} = 
	\mathcal W^{[0]} \mathcal A_0^{[-2]}\,,\qquad
	\frac1\epsilon\,
	\mathcal A_1^{[-1,-1]} = 
	\mathcal W^{[1]} \mathcal A_0^{[-2]}\,.
\end{equation}

\subsection{Factorization in the soft limit}
\label{ssec:soft}

In the limit
\begin{equation}\label{softLIMIT}
	k \sim  \mathcal O(\lambda) \,,\qquad \lambda\to0\,,
\end{equation}
the one-loop five-point amplitude $\mathcal A_1^{\mu\nu}$ must factorize according to the Weinberg soft graviton theorem \cite{Weinberg:1964ew} as
\begin{equation}\label{Fmunu}
	\mathcal F^{\mu\nu} = \sqrt{8\pi G}\, \sum_{n=1}^4 \frac{p_n^\mu p_n^\nu}{p_n\cdot k} \sim \mathcal O(\lambda^{-1})
\end{equation}
times the one-loop four-point amplitude $\mathcal A_1^{(4)}$,
\begin{equation}\label{Weinberg64factoriz}
	\mathcal A_1 = \mathcal F\, \mathcal A_1^{(4)} + \mathcal O(\lambda^0)\,,
\end{equation}
where  $\mathcal{F} = \varepsilon_\mu \mathcal{F}^{\mu\nu} \varepsilon_\nu$ in analogy with \eqref{contrA1}.
The limit \eqref{softLIMIT} is best taken after performing the decomposition in Eq.~\eqref{q12decomp} and following, introducing also the exchanged momentum
\begin{equation}\label{}
	q = \frac12(q_1-q_2)\,,
\end{equation}
whose decomposition reads
\begin{equation}\label{qdecomp}
	q^\mu = -\frac{\omega_1}{2} \check u_1^\mu + \frac{\omega_2}{2} \check u_2^\mu + q_\perp^\mu\,.
\end{equation}
The limit \eqref{softLIMIT} should then be understood for fixed $u_1$, $u_2$ and $q_\perp^\mu$, so that
\begin{equation}\label{}
	u_{1,2}\sim \mathcal O(\lambda^0)\,,\qquad
	q_\perp \sim \mathcal O(\lambda^0)\,,\qquad
	y \sim \mathcal O(\lambda^0)\,,\qquad
		\omega_{1,2} \sim \mathcal O(\lambda)
\end{equation}
and to leading order
\begin{equation}\label{}
	q_{1,2}^2 =q^2  + \mathcal O(\lambda)= q^2_\perp + \mathcal O(\lambda)\,,
	\qquad
	q_2^2-q_1^2 = 2 k\cdot q + \mathcal O(\lambda^2) = 2 k_\perp \cdot q_\perp +  \mathcal O(\lambda^2)\,.
\end{equation}
To leading order in this limit, the five-point kinematics \eqref{parq} reduces to the four-point one introduced in Ref.~\cite{Parra-Martinez:2020dzs},
\begin{align}\label{parq4}
	&p_1^\mu=-\bar m_1 u_1^\mu+q_{\perp}^\mu/2 \,,\qquad  p_4^\mu = \bar m_1 u_1^{\mu} + q_{\perp}^\mu/2 \\
	&p_2^\mu=-\bar m_2 u_2^{\mu} - q_{\perp}^\mu/2 \,,\qquad  p_3^\mu = \bar m_2 u_2^{\mu} - q_{\perp}^\mu/2 
\end{align}
with 
\begin{equation}\label{barmq11224}
	u_{1,2}\cdot q_{\perp} = 0\,,\qquad
	\bar m_1^2 = m_1^2 + \frac {q_\perp^2} 4, \qquad \bar m_2^2 = m_2^2 + \frac {q_\perp^2} 4\,.
\end{equation}

Both sides of the factorization \eqref{Weinberg64factoriz} should be expanded in the near-forward limit \eqref{hbarscaling} in order to be applied to our results.
One finds the following near-forward limit for the Weinberg factor \cite{DiVecchia:2021ndb},
\begin{equation}\label{Fexpansion}
	\mathcal F = \mathcal F^{[0]}  + \mathcal O(\hbar^2)
	\,,\qquad
	\mathcal F^{[0]} 
	=
	\mathcal O_\alpha q^\alpha
\end{equation}
with
\begin{equation}\label{}
	\mathcal O_\alpha
	=
	\sqrt{8\pi G}\,
	\frac{ ( \omega_1 (u_2\cdot\varepsilon)- \omega_2 (u_1\cdot\varepsilon )) (2 \omega_1\omega_2 \varepsilon_\alpha + k_\alpha \omega_2 (u_1\cdot \varepsilon)+ k_\alpha \omega_1 (u_2\cdot \varepsilon))}{\omega_1^2\omega_2^2}\,.
\end{equation}
Note the absence of $\mathcal O(\hbar^{1})$ corrections in \eqref{Fexpansion}.
For the other ingredient of \eqref{Weinberg64factoriz}, the one-loop four-point amplitude, one has instead \cite{Cheung:2018wkq,KoemansCollado:2019ggb,Cristofoli:2020uzm}
\begin{equation}\label{}
	\mathcal A_1^{(4)} = \mathcal A_1^{(4)[-2]} + \mathcal A_1^{(4)[-1]} + \mathcal O(\hbar^0)
\end{equation}
with the $\mathcal O(\hbar^{-2})$ term $\mathcal A_1^{(4)[-2]}=\tfrac1{\epsilon}\mathcal A_1^{(4)[-2,-1]}+\mathcal A_1^{(4)[-2,0]}+\mathcal O(\epsilon)$ given by 
\begin{equation}\label{}
	\mathcal A_1^{(4)[-2]} = 
	\bar{\mu}^{2\epsilon}\,
	\frac{i  32 \pi  G^2 \bar m_1^3 \bar m_2^3   \left(y^2-\tfrac{1}{2-2 \epsilon }\right)^2}{\sqrt{y^2-1}} \frac{e^{\gamma_E \epsilon} \Gamma (\epsilon +1) \Gamma (-\epsilon )^2}{(q^2)^{1+\epsilon}\Gamma (-2 \epsilon )}
\end{equation}
for generic $\epsilon$, while we will only need the $\mathcal O(\hbar^{-1})$ term $\mathcal A_1^{(4)[-1]}=\mathcal A_1^{(4)[-1,0]}+\mathcal O(\epsilon)$ to leading order in $\epsilon$,
\begin{equation}\label{triangles}
	\mathcal A_1^{(4)[-1,0]} = \frac{6 \pi^2 G^2  \bar m_1^2 \bar m_2^2 \left(5 y^2-1\right) (\bar m_1+\bar m_2)}{q} \,.
\end{equation}
Eq.~\eqref{Weinberg64factoriz} then translates into the following relations
\begin{equation}\label{W64m2m1}
	\mathcal A_1^{[-2]} = \mathcal F^{[0]} \mathcal A_1^{(4)[-2]}+\mathcal O(\lambda^0)\,,\qquad
	\mathcal A_1^{[-1]} = \mathcal F^{[0]} \mathcal A_1^{(4)[-1]}+\mathcal O(\lambda^0)\,,
\end{equation}
which can be also expanded for small $\epsilon$. 

Of course, the soft limit of the $1/\epsilon$ terms, 
\begin{equation}\label{am2m1W64}
	\frac{1}{\epsilon}\,
	\mathcal A_1^{[-2,-1]} = \mathcal F^{[0]} \mathcal A_1^{(4)[-2,-1]}+\mathcal O(\lambda^0)\,,\qquad
	\frac{1}{\epsilon}\,\mathcal A_1^{[-1,-1]} = \mathcal O(\lambda^0)
	\,,
\end{equation}
is already captured by the exponentiation of infrared divergences discussed in Subsection~\ref{ssec:IR}. 
If one considers the soft limit of Eq.~\eqref{am2m1W65} and takes into account that the tree-level amplitude obeys the Weinberg theorem, $\mathcal A_0^{[-2]} = \mathcal F^{[0]} \mathcal A_0^{(4)} + \mathcal O(\lambda^0)$, then \eqref{am2m1W65} becomes equivalent to \eqref{am2m1W64}, because $\mathcal A_1^{(4)}$ also obeys the factorization of infrared divergences. For the $\mathcal O(\hbar^{-2})$ term, this dictates $\tfrac1{\epsilon}\mathcal A_1^{(4)[-2,-1]} = \mathcal W^{[0]} \mathcal A_0^{(4)}$, i.e. (see e.g.~Eq.~(3.52) of \cite{Heissenberg:2021tzo})
\begin{equation}\label{}
	\frac{1}{\epsilon}\,
	\mathcal A_1^{(4)[-2,-1]} = \frac{64 G^2 \bar m_1^3 \bar m_2^3}{\epsilon q^2} \left(
	y^2-\tfrac12
	\right)^2
	\frac{-i \pi}{\sqrt{y^2-1}}\,.
\end{equation}
For the $\mathcal O(\hbar^{-1})$ terms in \eqref{am2m1W64}, the absence of a Weinberg pole $\sim 1/\lambda$ is also ensured by \eqref{am2m1W65} because $\mathcal W^{[1]}$ carries an extra power of $\lambda$.

Constraints on the soft limit of $\mathcal A_1$ independent of the exponentiation of infrared divergences instead involve the finite parts,
\begin{equation}\label{}
	\mathcal A_1^{[-2,0]} = \mathcal F^{[0]} \mathcal A_1^{(4)[-2,0]}+\mathcal O(\lambda^0)\,,\qquad
	\mathcal A_1^{[-1,0]} = \mathcal F^{[0]} \mathcal A_1^{(4)[-1,0]} +\mathcal O(\lambda^0)\,.
\end{equation}
We have checked that our results are consistent with these constraints.
It is instructive to see how they translate to impact-parameter space, by letting
\begin{equation}\label{FT4}
\operatorname{FT}[\mathcal A^{(4)}] = 	\tilde{\mathcal{A}}^{(4)}(b) = \int \mathcal A^{(4)}(q)\, 2\pi\delta(2\bar m_1u_1\cdot q) 2\pi\delta(2\bar m_2u_2\cdot q)  e^{ib\cdot q}\frac{d^Dq}{(2\pi)^D}\,,
\end{equation}
where the $2\to2$ amplitude (see~\eqref{A04pt}, \eqref{analytic} for the tree level) obeys the eikonal exponentiation \cite{KoemansCollado:2019ggb,Cristofoli:2020uzm}
\begin{equation}\label{b4pt}
	\tilde{\mathcal{A}}^{(4)}_0 = 2\delta_0\,,\qquad
	\tilde{\mathcal{A}}^{(4)}_1 = i\,\frac{(2\delta_0)^2}{2} + 2\delta_1\,.
\end{equation}
Since we work to leading order in the soft limit, we can apply the same Fourier transform \eqref{FT4} to the $2\to3$ amplitude as well, finding that \eqref{W64m2m1} translates to
\begin{equation}\label{bsoft1loop}
	\tilde{\mathcal{A}}_1^{[-2]} = 2\delta_0 \, \mathcal O^\alpha Q^\text{1PM}_\alpha +\mathcal O(\lambda^0)\,,\qquad
	\tilde{\mathcal{A}}_1^{[-1]} = -i\mathcal O^\alpha Q^\text{2PM}_\alpha +\mathcal O(\lambda^0)\,,
\end{equation}
were we have used that $\operatorname{FT}[q^\alpha(\,\cdots)]=-i\partial_b^\alpha\operatorname{FT}[\,\cdots]$ and the relation between the impulse and the eikonal phase up to 2PM,
\begin{equation}\label{impulse}
	Q_\alpha = \frac{\partial 2\delta}{\partial b^\alpha}\,,\qquad
	2\delta= 2\delta_0 + 2\delta_1+\mathcal O(G^3)\,,\qquad
	Q= Q^\text{1PM}+ Q^\text{2PM}+\mathcal O(G^3)\,.
\end{equation}
Of course the soft theorem also holds for the tree-level amplitude, and one has
\begin{equation}\label{bsofttree}
	{\mathcal{A}}_0^{[-2]} = \mathcal O^\alpha q_\alpha \mathcal A_0^{(4)}+\mathcal O(\lambda^0)\,,
	\qquad
	\tilde{\mathcal{A}}_0^{[-2]} = -i \mathcal O^\alpha Q^\text{1PM}_\alpha+\mathcal O(\lambda^0)\,.
\end{equation}
Combining \eqref{bsoft1loop} and \eqref{bsofttree} provides a check of the leading inelastic exponentiation of the one-loop level amplitude in $b$-space, to first order in the soft limit
\begin{equation}\label{}
	\tilde{\mathcal{A}}_1^{[-2]} = 2i\delta_0 \, \tilde{\mathcal{A}}_0^{[-2]} +\mathcal O(\lambda^0)\,,
\end{equation}
whereby the ``superclassical'' term of the one-loop inelastic amplitude factorizes in terms of the elastic tree-level amplitude times the inelastic one in $b$-space. 
Moreover, the relations \eqref{bsoft1loop} can be seen as a manifestation order by order in $G$ of the non-perturbative pattern discussed in \cite{DiVecchia:2022owy,DiVecchia:2022nna} according to which the soft dressing governing the soft theorem/memory effect for processes with generic deflections can be obtained from the Weinberg factor \eqref{Fmunu} by replacing  the perturbative momentum transfer $q^\alpha$ with the classical impulse $Q^\alpha$ given by \eqref{impulse}.

\subsection{Imaginary parts and unitarity}
\label{ssec:unitarity}

The ``superclassical'' contributions\footnote{The discussion of real and imaginary parts of the amplitude should be performed by either stripping off the polarization tensor, which has both real and imaginary parts in our case, or by keeping it while disregarding factors of ``$i$'' arising from it.} $\mathcal A_1^{\mu\nu[-2,-1]}$, $\mathcal A_1^{\mu\nu[-2,0]}$ are purely imaginary and, by unitarity, they must correspond to appropriate intermediate states for the $2\to3$ process under consideration. We find that they are equal to the the sum of two processes where one ``cuts'' two intermediate massive states, which we may term ``$S$-channel'' by analogy with the situation at four points,
\begin{equation}\label{s-channel-cuts}
	\bar\mu^{2\epsilon}\left[
	\frac{1}{\epsilon}\,\mathcal A_1^{\mu\nu[-2,-1]}+ \mathcal A_1^{\mu\nu[-2,0]} \right] = 
	\frac{i}{2} \
	\begin{gathered}
		\begin{tikzpicture}
			\path [draw, ultra thick, blue] (-4,2)--(-.3,2);
			\path [draw, ultra thick, green!60!black] (-4,1)--(-.3,1);
			\path [draw, red] (-1,1.5)--(-.32,1.5);
			\filldraw[black!20!white, thick] (-3,1.5) ellipse (.5 and .8);
			\draw[thick] (-3,1.5) ellipse (.5 and .8);
			\filldraw[black!20!white, thick] (-1.3,1.5) ellipse (.5 and .8);
			\draw[thick] (-1.3,1.5) ellipse (.5 and .8);
		\end{tikzpicture}
	\end{gathered}
	+
	\frac{i}{2} \
	\begin{gathered}
		\begin{tikzpicture}
			\path [draw, ultra thick, blue] (-4,2)--(-.3,2);
			\path [draw, ultra thick, green!60!black] (-4,1)--(-.3,1);
			\path [draw, red] (-3,1.5)--(-2.1,1.5);
			\filldraw[black!20!white, thick] (-3,1.5) ellipse (.5 and .8);
			\draw[thick] (-3,1.5) ellipse (.5 and .8);
			\filldraw[black!20!white, thick] (-1.3,1.5) ellipse (.5 and .8);
			\draw[thick] (-1.3,1.5) ellipse (.5 and .8);
		\end{tikzpicture}
	\end{gathered}
\end{equation}
The first process on the right-hand side is obtained by gluing together a tree-level $2\to2$ amplitude $\mathcal A_0^{(4)}$ involving four massive states \eqref{A04pt} and a tree-level $2\to3$ amplitude $\mathcal A_0$ for the inelastic process under consideration \eqref{A05pt}.
The second process formally corresponds to a $2\to3$ amplitude glued together with a partially disconnected $3\to3$ one, in which the graviton line simply ``passes through'' the second blob, so that in practice it corresponds to gluing together $\mathcal A_0^{(4)}$ and $\mathcal A_0$ in the opposite order. Equivalently, it can be obtained from the first one by applying the permutation $\sigma_2\sigma_3$, which flips $\omega_1$, $\omega_2$ and leaves $y$ unaltered.

We have checked Eq.~\eqref{s-channel-cuts} in two ways. First, by leaving the signs of the analytic continuations arbitrary as in \eqref{analyticcontinuationLOGs}, i.e.~without imposing \eqref{qSqC}, one sees that the left-hand side of \eqref{s-channel-cuts} is only sensitive to $q_I$, the sign of the analytic continuation of $y$, which is the invariant associated to propagation in the $S$-channel. More precisely, denoting by $f(q_I,q_O,q_A)$ this generalized version of the amplitude obtained by leaving the three signs of the analytic continuations arbitrary, the left-hand side of \eqref{s-channel-cuts} coincides with the $S$-channel discontinuity 
\begin{equation}\label{s-channel-discontinuity}
	\begin{aligned}
		\operatorname{Disc}_S f 
		= \frac{1}{8} \, [f(+1,+1,+1)+f(+1,+1,-1)+f(+1,-1,+1) &+f(+1,-1,-1)\\
		-f(-1,+1,+1)-f(-1,+1,-1)-f(-1,-1,+1) &-f(-1,-1,-1)
		].
	\end{aligned}
\end{equation}
Second, we have explicitly constructed the integrand for the right-hand side of \eqref{s-channel-cuts} by gluing together $\mathcal A_0^{(4)}$ and $\mathcal A_0$ (for completeness, we provide their explicit expressions in Appendix~\ref{appendix:treelevel}) in the classical limit and performed the integration over phase space via reverse unitarity \cite{Anastasiou:2002qz,Anastasiou:2002yz,Anastasiou:2003yy,Anastasiou:2015yha,Herrmann:2021tct,Herrmann:2021lqe}. This amounts to treating the Lorentz-invariant phase space delta functions like formal propagators, performing the IBP reduction while dropping all integrals that do not possess the cut \eqref{s-channel-cuts}, and lastly substituting the master integrals with $2$ times their imaginary parts obtained by applying $\operatorname{Disc}_S$ as defined by \eqref{s-channel-discontinuity} (i.e.~the imaginary parts associated to their $S$-channel discontinuities).

The purely imaginary term $\mathcal A_1^{\mu\nu[-1,-1]}$ and the imaginary part of $\mathcal A_1^{\mu\nu[-1,0]}$ arise instead due to intermediate processes whereby one cuts a massive line and a graviton line. Since these cuts are built using the ``Compton'' amplitude \eqref{ComptonAmplitude} involving two massive states and two gravitons, together with the tree-level $2\to3$ amplitude $\mathcal A_0$, we may term these ``$C$-channel'' cuts,
\begin{equation}\label{C-channel-cuts}
		\bar\mu^{2\epsilon}\left[
	\frac{1}{\epsilon}\mathcal A_1^{\mu\nu[-1,-1]} + i \operatorname{Im}\mathcal A_1^{\mu\nu[-1,0]}\right] = 
	\frac{i}{2} \
	\begin{gathered}
		\begin{tikzpicture}
			\path [draw, ultra thick, blue] (-4,2)--(-.3,2);
			\path [draw, ultra thick, green!60!black] (-4,1)--(-2.1,1);
			\path [draw, red] (-3,1.5)--(-.3,1.5);
			\filldraw[black!20!white, thick] (-3,1.5) ellipse (.5 and .8);
			\draw[thick] (-3,1.5) ellipse (.5 and .8);
			\filldraw[black!20!white, thick] (-1.3,1.75) ellipse (.45 and .55);
			\draw[thick] (-1.3,1.75) ellipse (.45 and .55);
		\end{tikzpicture}
	\end{gathered}
	+
	\frac{i}{2} \
	\begin{gathered}
		\begin{tikzpicture}
			\path [draw, ultra thick, blue] (-4,2)--(-2.1,2);
			\path [draw, ultra thick, green!60!black] (-4,1)--(-.3,1);
			\path [draw, red] (-3,1.5)--(-.3,1.5);
			\filldraw[black!20!white, thick] (-3,1.5) ellipse (.5 and .8);
			\draw[thick] (-3,1.5) ellipse (.5 and .8);
			\filldraw[black!20!white, thick] (-1.3,1.25) ellipse (.45 and .55);
			\draw[thick] (-1.3,1.25) ellipse (.45 and .55);
		\end{tikzpicture}
	\end{gathered}
\end{equation}
Once again these processes can be also thought of as $2\to3$ amplitudes glued with partially disconnected $3\to3$ ones.
We have checked \eqref{C-channel-cuts} in the same two ways as for \eqref{s-channel-cuts}, both by calculating the discontinuity of the  $f(q_I,q_O,q_A)$ with respect to $q_O$ (and separately $q_A$), and by building the integrand for the cuts and evaluating it via reverse unitarity.\footnote{For this check, we find that the trace condition \eqref{trace=0} plays an important role.}

Combining \eqref{s-channel-cuts} with \eqref{C-channel-cuts}, we reconstruct the complete unitarity relation for the one-loop amplitude,
\begin{equation}\label{fullUnitarity}
	\begin{split}
	2\operatorname{Im}
	\begin{gathered}
		\begin{tikzpicture}
			\path [draw, ultra thick, blue] (-4.5,2.2)--(-1.5,2.2);
			\path [draw, ultra thick, green!60!black] (-4.5,.8)--(-1.5,.8);
			\path [draw, red] (-3,1.5)--(-1.5,1.5);
			\filldraw[black!20!white, thick] (-3,1.5) ellipse (.7 and 1);
			\draw[thick] (-3,1.5) ellipse (.7 and 1);
			\filldraw[white, thick] (-3,1.5) ellipse (.3 and .4);
			\filldraw[pattern=none, thick] (-3,1.5) ellipse (.3 and .4);
		\end{tikzpicture}
	\end{gathered}
&=\
\begin{gathered}
	\begin{tikzpicture}
		\path [draw, ultra thick, blue] (-4,2)--(-.3,2);
		\path [draw, ultra thick, green!60!black] (-4,1)--(-.3,1);
		\path [draw, red] (-1,1.5)--(-.32,1.5);
		\filldraw[black!20!white, thick] (-3,1.5) ellipse (.5 and .8);
		\draw[thick] (-3,1.5) ellipse (.5 and .8);
		\filldraw[black!20!white, thick] (-1.3,1.5) ellipse (.5 and .8);
		\draw[thick] (-1.3,1.5) ellipse (.5 and .8);
	\end{tikzpicture}
\end{gathered}
+
 \
\begin{gathered}
	\begin{tikzpicture}
		\path [draw, ultra thick, blue] (-4,2)--(-.3,2);
		\path [draw, ultra thick, green!60!black] (-4,1)--(-.3,1);
		\path [draw, red] (-3,1.5)--(-2.1,1.5);
		\filldraw[black!20!white, thick] (-3,1.5) ellipse (.5 and .8);
		\draw[thick] (-3,1.5) ellipse (.5 and .8);
		\filldraw[black!20!white, thick] (-1.3,1.5) ellipse (.5 and .8);
		\draw[thick] (-1.3,1.5) ellipse (.5 and .8);
	\end{tikzpicture}
\end{gathered}
\\
&
+\
\begin{gathered}
	\begin{tikzpicture}
		\path [draw, ultra thick, blue] (-4,2)--(-.3,2);
		\path [draw, ultra thick, green!60!black] (-4,1)--(-2.1,1);
		\path [draw, red] (-3,1.5)--(-.3,1.5);
		\filldraw[black!20!white, thick] (-3,1.5) ellipse (.5 and .8);
		\draw[thick] (-3,1.5) ellipse (.5 and .8);
		\filldraw[black!20!white, thick] (-1.3,1.75) ellipse (.45 and .55);
		\draw[thick] (-1.3,1.75) ellipse (.45 and .55);
	\end{tikzpicture}
\end{gathered}
+
\
\begin{gathered}
	\begin{tikzpicture}
		\path [draw, ultra thick, blue] (-4,2)--(-2.1,2);
		\path [draw, ultra thick, green!60!black] (-4,1)--(-.3,1);
		\path [draw, red] (-3,1.5)--(-.3,1.5);
		\filldraw[black!20!white, thick] (-3,1.5) ellipse (.5 and .8);
		\draw[thick] (-3,1.5) ellipse (.5 and .8);
		\filldraw[black!20!white, thick] (-1.3,1.25) ellipse (.45 and .55);
		\draw[thick] (-1.3,1.25) ellipse (.45 and .55);
	\end{tikzpicture}
\end{gathered}
\end{split}
\end{equation}

\subsection{Removing superclassical iterations}
\label{ssec:iterations}
 
We follow Ref.~\cite{Damgaard:2021ipf} and consider the operator $N$ linked to the $S$-matrix by
\begin{equation}\label{SdefN}
	S = 1+i T = e^{iN}\,,\qquad N = -i \log(1+iT) = T - i\, \frac{T^2}{2}+\cdots\,.
\end{equation}
As usual, we define their matrix elements by stripping a momentum-conserving delta function,
\begin{equation}\label{}
	\langle \beta | T |\alpha\rangle= (2\pi)^D \delta^{(D)}(P_\alpha+P_\beta) \mathcal A_{\alpha\to\beta}\,,
\qquad
	\langle \beta | N |\alpha\rangle= (2\pi)^D \delta^{(D)}(P_\alpha+P_\beta) \mathcal B_{\alpha\to\beta}\,,
\end{equation}
where $\mathcal A_{\alpha\to\beta}$ are the conventional scattering amplitudes.
From \eqref{SdefN}, one trivially obtains 
\begin{equation}\label{}
	\mathcal B_0^{\mu\nu} = \mathcal A_0^{\mu\nu}
\end{equation}
at tree level. Going to the next order and
inserting a complete set of free intermediate states to resolve the terms involving $T^2$, one  finds
\begin{equation}\label{subtractionsN}
	\begin{split}
	\mathcal B_1^{\mu\nu}
	=
	\mathcal A_1^{\mu\nu} 
	&
	-
	\frac{i}{2}\
	\begin{gathered}
		\begin{tikzpicture}
			\path [draw, ultra thick, blue] (-4,2)--(-.3,2);
			\path [draw, ultra thick, green!60!black] (-4,1)--(-.3,1);
			\path [draw, red] (-1,1.5)--(-.32,1.5);
			\filldraw[black!20!white, thick] (-3,1.5) ellipse (.5 and .8);
			\draw[thick] (-3,1.5) ellipse (.5 and .8);
			\filldraw[black!20!white, thick] (-1.3,1.5) ellipse (.5 and .8);
			\draw[thick] (-1.3,1.5) ellipse (.5 and .8);
		\end{tikzpicture}
	\end{gathered}
	-
	\frac{i}{2}
	\
	\begin{gathered}
		\begin{tikzpicture}
			\path [draw, ultra thick, blue] (-4,2)--(-.3,2);
			\path [draw, ultra thick, green!60!black] (-4,1)--(-.3,1);
			\path [draw, red] (-3,1.5)--(-2.1,1.5);
			\filldraw[black!20!white, thick] (-3,1.5) ellipse (.5 and .8);
			\draw[thick] (-3,1.5) ellipse (.5 and .8);
			\filldraw[black!20!white, thick] (-1.3,1.5) ellipse (.5 and .8);
			\draw[thick] (-1.3,1.5) ellipse (.5 and .8);
		\end{tikzpicture}
	\end{gathered}
	\\
	&
	-
	\frac{i}{2}\
	\begin{gathered}
		\begin{tikzpicture}
			\path [draw, ultra thick, blue] (-4,2)--(-.3,2);
			\path [draw, ultra thick, green!60!black] (-4,1)--(-2.1,1);
			\path [draw, red] (-3,1.5)--(-.3,1.5);
			\filldraw[black!20!white, thick] (-3,1.5) ellipse (.5 and .8);
			\draw[thick] (-3,1.5) ellipse (.5 and .8);
			\filldraw[black!20!white, thick] (-1.3,1.75) ellipse (.45 and .55);
			\draw[thick] (-1.3,1.75) ellipse (.45 and .55);
		\end{tikzpicture}
	\end{gathered}
	-
	\frac{i}{2}
	\
	\begin{gathered}
		\begin{tikzpicture}
			\path [draw, ultra thick, blue] (-4,2)--(-2.1,2);
			\path [draw, ultra thick, green!60!black] (-4,1)--(-.3,1);
			\path [draw, red] (-3,1.5)--(-.3,1.5);
			\filldraw[black!20!white, thick] (-3,1.5) ellipse (.5 and .8);
			\draw[thick] (-3,1.5) ellipse (.5 and .8);
			\filldraw[black!20!white, thick] (-1.3,1.25) ellipse (.45 and .55);
			\draw[thick] (-1.3,1.25) ellipse (.45 and .55);
		\end{tikzpicture}
	\end{gathered}
\end{split}
\end{equation}
In view of the unitarity relation \eqref{fullUnitarity}, all imaginary parts of $\mathcal A_1^{\mu\nu}$ cancel out and one is left with the real and IR-finite result
\begin{equation}\label{B1}
\mathcal B_1^{\mu\nu} = \operatorname{Re}\mathcal A_1^{\mu\nu} = \operatorname{Re}\mathcal A_1^{\mu\nu[-1,0]} + \mathcal O(\epsilon) + \mathcal O(\hbar^0) \,.
\end{equation}
In fact, the subtractions in the first line of \eqref{subtractionsN} are enough in order to remove all $\mathcal O(\hbar^{-2})$ ``superclassical'' terms, while as already discussed the ones in the second line are $\mathcal O(\hbar^{-1})$. Both types of subtractions involve imaginary infrared divergences, the ones in the first line being associated to $\mathcal W^{[0]}$, i.e.~to soft-graviton exchanges between massive lines, and the ones in the second line being associated to $\mathcal W^{[1]}$, i.e.~soft-graviton exchanges between the graviton and an outgoing massive line, as also suggested by the respective figures.
Letting $\mathcal B_1 = \varepsilon_\mu\mathcal B_1^{\mu\nu}\varepsilon_\nu$, as already mentioned, $\mathcal B_1/\mathcal N_4$ has uniform transcendental weight 2 and takes the form of $\pi^2$  multiplying a rational function of the invariants.

Let us conclude this section by commenting on the parity properties of $\mathcal B_1$
under the transformation\footnote{As already mentioned, also $w_{1,2}\mapsto-w_{1,2}$ corresponds to changing the sign of $\omega_{1,2}$, but the transformation in \eqref{flipomega} is the one that leaves $\sqrt{\omega_{1,2}^2+q_{2,1}^2}=\tfrac{q_{2,1}}{2}\left(w_{1,2}+\tfrac{1}{w_{1,2}}\right)$ invariant.}
\begin{equation}\label{flipomega}
	\omega_{1,2} \mapsto -\omega_{1,2}\,,\qquad
	w_{1,2} \mapsto \frac{1}{w_{1,2}}\,.
\end{equation}
 We find that
 \begin{equation}\label{}
 	\mathcal B_{1} = \mathcal B_{1E} + \mathcal B_{1O}\,, 
 \end{equation}
with $\mathcal B_{1E}$ (resp.~$\mathcal B_{1O}$) even (odd) under \eqref{flipomega}.
In particular, the odd piece is equal to an $x$-dependent function times $i\pi$ times the coefficient of the $\hbar^{-1}\epsilon^{-1}$ pole
\begin{equation}\label{}
	\mathcal B_{1O}
	=
	\frac{2 x^4 \left(x^2-3\right)}{\left(x^2-1\right)^3}
	(i \pi) \mathcal A_1^{[-1,-1]}
	=
	\left(
	1-
	\frac{y \left(y^2-\frac{3}{2}\right)}{\left(y^2-1\right)^{3/2}}\right)
	(i \pi) \mathcal A_1^{[-1,-1]}
	\,.
\end{equation}
Time-reversal-odd terms in the finite real part thus arise from the analytic continuation of logarithms left behind by the $1/\epsilon$ in the imaginary part. This mechanism is highly reminiscent of how radiation reaction enters the eikonal phase at two loops \cite{DiVecchia:2021ndb,DiVecchia:2021bdo,Alessio:2022kwv,DiVecchia:2022owy,DiVecchia:2022nna} (see also \cite{SergolaTalk,Donal&Co}).

\section{Gravitational Field, Spectrum and Waveform}
\label{sec:waveform}

Following Refs.~\cite{Kosower:2018adc,Cristofoli:2021vyo,Cristofoli:2021jas}, let us model the initial state of the collision by 
\begin{equation}\label{}
	| \text{in} \rangle = |1\rangle \otimes |2\rangle\,,
\end{equation}
with
\begin{align}\label{}
	|1\rangle &=\int 2\pi\delta(p_1^2+m_1^2) \theta(-p_1^0) \frac{d^Dp_1}{(2\pi)^D}
\,\varphi_1(-p_1) \,e^{ib_1\cdot p_1}|-p_1\rangle\\
	|2\rangle &=
	\int 2\pi\delta(p_2^2+m_2^2) \theta(-p_2^0) \frac{d^Dp_2}{(2\pi)^D}\,\varphi_2(-p_2)\,e^{ib_2\cdot p_2}|-p_2\rangle
\end{align}
in terms of wavepackets $\varphi_{1,2}$ and impact parameters $b_1$, $b_2$.
We can then take the expectation value of the graviton field
\begin{equation}\label{}
	H_{\mu\nu}(x) = \int_k \left[
	e^{ik\cdot x} a_{\mu\nu}(k)
	+
	e^{-ik\cdot x} a^{\dagger}_{\mu\nu}(k)
	\right]
	,\qquad
	\int_k=
	 \int 2\pi\delta(k^2) \theta(k^0)
	\frac{d^Dk}{(2\pi)^D}\,,
\end{equation}
in the state obtained by applying the $S$ matrix \eqref{SdefN},
\begin{equation}\label{}
	|\text{out}\rangle  = S |\text{in} \rangle\,.
\end{equation}
We denote this expectation by (``c.c'' stands for ``complex conjugate'')
\begin{equation}\label{}
	\frac{g_{\mu\nu}(x)-\eta_{\mu\nu}}{\sqrt{32\pi G}}
	=
	h_{\mu\nu}(x) = \langle \text{out} | H_{\mu\nu}(x) |\text{out}\rangle
	=
	\int_k
	e^{ik\cdot x}  \langle\text{out}|a_{\mu\nu}(k)|\text{out}\rangle
	+(\text{c.c.})\,.
\end{equation}
Defining the Fourier transform by the generalization of \eqref{FT4} to the $2\to3$ kinematics,
\begin{equation}\label{FT5}
	\begin{split}
	\operatorname{FT}\left[
	f^{\mu\nu}
	\right]
	=
	\tilde{f}^{\mu\nu}
	&=
	\int \frac{d^Dq_1}{(2\pi)^D}\frac{d^Dq_2}{(2\pi)^D}\,(2\pi)^D\delta^{(D)}(q_1+q_2+k) \\
	&\times
	2\pi\delta(2\bar m_1 u_1\cdot q_1)2\pi\delta(2\bar m_2 u_2\cdot q_2)
	e^{ib_1\cdot q_1+ib_2\cdot q_2} f^{\mu\nu}
	\end{split}
\end{equation}
and introducing a shorthand notation for the wavepacket average
\begin{equation}\label{}
	\langle f \rangle = \int \prod_{j=1,2} 2\pi\delta(\bar p_j^2+ \bar m_j^2) \theta(\bar p_j^0) \frac{d^D \bar p_j}{(2\pi)^D}
	\,\varphi_j(\bar p_j-\tfrac 12 q_j) \varphi_j^\ast(\bar p_j+\tfrac 12 q_j)  f\,,
\end{equation}
we find
\begin{equation}\label{hmunuWmunu}
	h_{\mu\nu}(x)
	=
	\int_k\left[
	e^{ik\cdot x}\,
	i
	\operatorname{FT}
	\langle\, W_{\mu\nu} \rangle
	\right]
	+
	\text{c.c.}\,,\qquad
	W^{\mu\nu} = W_0^{\mu\nu} + W_1^{\mu\nu} 
\end{equation}
where 
\begin{equation}\label{}
	\begin{split}
	W_{0}^{\mu\nu} 
	&= \mathcal B_{0}^{\mu\nu}  =\mathcal A_{0}^{\mu\nu} \,,\\
	\label{W1B1i2i2}
	W_{1}^{\mu\nu}
	&=
	\mathcal B_{1}^{\mu\nu} 
	+
		\frac{i}{2}\
	\begin{gathered}
		\begin{tikzpicture}
			\path [draw, ultra thick, blue] (-4,2)--(-.3,2);
			\path [draw, ultra thick, green!60!black] (-4,1)--(-2.1,1);
			\path [draw, red] (-3,1.5)--(-.3,1.5);
			\filldraw[black!20!white, thick] (-3,1.5) ellipse (.5 and .8);
			\draw[thick] (-3,1.5) ellipse (.5 and .8);
			\filldraw[black!20!white, thick] (-1.3,1.75) ellipse (.45 and .55);
			\draw[thick] (-1.3,1.75) ellipse (.45 and .55);
		\end{tikzpicture}
	\end{gathered}
	+
	\frac{i}{2}
	\
	\begin{gathered}
		\begin{tikzpicture}
			\path [draw, ultra thick, blue] (-4,2)--(-2.1,2);
			\path [draw, ultra thick, green!60!black] (-4,1)--(-.3,1);
			\path [draw, red] (-3,1.5)--(-.3,1.5);
			\filldraw[black!20!white, thick] (-3,1.5) ellipse (.5 and .8);
			\draw[thick] (-3,1.5) ellipse (.5 and .8);
			\filldraw[black!20!white, thick] (-1.3,1.25) ellipse (.45 and .55);
			\draw[thick] (-1.3,1.25) ellipse (.45 and .55);
		\end{tikzpicture}
	\end{gathered}
\end{split}
\end{equation}
Notably, although $\mathcal B_0$ and $\mathcal B_1$ are real and finite, the two infrared divergent $C$-channel cuts have ``reappeared'' in the loop-level result for the KMOC kernel $W_1^{\mu\nu}$ in \eqref{W1B1i2i2}. This comes about because, to this order, using $S = e^{iN} = 1+iN-\tfrac12 N^2+\cdots$,
\begin{equation}\label{aaNaNNNaN}
	\langle\text{out}|a_{\mu\nu}(k)|\text{out}\rangle
	=
	i \langle\text{in}|a_{\mu\nu}(k) N|\text{in}\rangle
	- \frac{1}{2} \langle\text{in}|a_{\mu\nu}(k) N^2|\text{in}\rangle
	+
	\langle\text{in}|N a_{\mu\nu}(k) N|\text{in}\rangle + \cdots\,.
\end{equation}
Inserting a complete set of states, we see that at one loop the term $\langle\text{in}|a_{\mu\nu}(k) N^2|\text{in}\rangle$ includes all cuts depicted in Table~\ref{tab:channels}, while $\langle\text{out}|N a_{\mu\nu}(k) N|\text{out}\rangle$ only contains the second $S$-channel cut. Thanks to the factor of $-\tfrac12$, this leads to a cross-cancellation of the $S$-channel cuts in \eqref{aaNaNNNaN}, which leaves behind the $C$-channel cuts as in \eqref{W1B1i2i2}.\footnote{In Ref.~\cite{Caron-Huot:2023vxl} it was pointed out that the leading-order cancellation of the superclassical part of the $S$-channel cuts also leaves behind a nontrivial classical part. See Refs.~\cite{Georgoudis:2023eke,Georgoudis:2023ozp} for its inclusion.}

We may rewrite $W_1^{\mu\nu}$ as follows\footnote{From now on, we drop the explicit wavepacket average, and neglect the difference between, say $\bar m_{1,2}$ and $m_{1,2}$, since superclassical terms have all been subtracted out.}
\begin{equation}\label{}
	W_{1}^{\mu\nu}
	=
	\mathcal B_1^{\mu\nu}
	-i
	\bar\mu^{2\epsilon}
	\frac{G}{\epsilon}\,(m_1\omega_1+m_2\omega_2)\mathcal B_0^{\mu\nu}
	+i
	\operatorname{Im}\mathcal A_1^{\mu\nu[-1,0]}
	+\mathcal O(\epsilon)
\end{equation}
where we have used that the infrared divergence is proportional to the tree-level result by \eqref{am2m1W65} and \eqref{W0W1}.
By exponentiating it, we may also rewrite $W^{\mu\nu}$ in the following way
\begin{equation}\label{Wmunu}
W^{\mu\nu} = e^{-i\frac{G}{\epsilon}(m_1\omega_1+m_2\omega_2)}
\left[
\mathcal B_0^{\mu\nu}
+
\mathcal B_1^{\mu\nu}
+i
\mathcal M_1^{\mu\nu}
\right]
+\mathcal O(\epsilon) + \mathcal O(G^{7/2})
\end{equation}
in terms of the infrared-finite object
\begin{equation}\label{M1munu}
	\mathcal M_1^{\mu\nu}
	=
	G(m_1\omega_1+m_2\omega_2)
	\mathcal B_0^{\mu\nu}
	\log\bar\mu^2
	+
	\operatorname{Im}\mathcal  A_1^{\mu\nu[-1,0]}
	\,.
\end{equation}
As usual, the factorization of an infrared-divergent scale has left behind the logarithm of an arbitrary scale, and one can verify that such $\log\bar\mu^2$ terms neatly combine with the leftover $\log q_{1,2}^2$ terms in $\operatorname{Im}\mathcal  A_1^{\mu\nu[-1,0]}$ to reconstruct logarithms of dimensionless quantities.

Let us now turn to the spectral rate,
\begin{equation}\label{}
	d\tilde\rho = |\tilde W|^2 \theta(k^0)\,2\pi\delta(k^2)\,\frac{d^4k}{(2\pi)^4}\,,
\end{equation}
where omitting the $\mu\nu$ indices stands for contraction according to
\begin{equation}\label{}
	|\tilde W|^2 = \tilde W^{\mu\nu} \left( \eta_{\mu\rho}\eta_{\nu\sigma} - \frac{1}{2}\,\eta_{\mu\nu}\eta_{\rho\sigma} \right) \tilde W^{\rho\sigma\ast}\,.
\end{equation}
Then it is clear that, although $W^{\mu\nu}$ in \eqref{Wmunu} has an IR-divergent phase, the spectrum is free from infrared divergences, since this overall phase cancels out, and retaining terms up to $\mathcal O(G^4)$,
\begin{equation}\label{}
	|\tilde W|^2 =  \tilde{\mathcal{B}}_0^\ast  \tilde{\mathcal{B}}_0 + (\tilde{\mathcal{B}}_0^\ast \, \tilde{\mathcal{B}}_1 + \tilde{\mathcal{B}}_1^\ast \, \tilde{\mathcal{B}}_0) -i (\tilde{\mathcal{B}}_0^\ast \, \tilde{\mathcal{M}}_1 - \tilde{\mathcal{M}}_1^\ast\, \tilde{\mathcal{B}}_0) + \mathcal O(G^5)\,.
\end{equation}
In principle this cancellation could leave behind ambiguities associated to the $\log\bar\mu^2$ terms in \eqref{Wmunu}, \eqref{M1munu}.
To see why this is not the case, let us denote by
\begin{equation}\label{}
	 E(k) = G( m_1 \omega_1 + m_2 \omega_2)
\end{equation} 
the combination appearing in the $\log \bar \mu^2$ terms, which is insensitive to the Fourier transform \eqref{FT5}.
Terms of type $\mathcal B_0  E \log\bar\mu^2$ in the imaginary part $\mathcal M_1$ of the waveform kernel enter the spectral rate via
\begin{align}\label{}
	\tilde{\mathcal{B}}_0^\ast \tilde{\mathcal{B}}_0 E(k) \log\bar\mu^2 - E(k)\tilde{\mathcal{B}}_0 \tilde{\mathcal{B}}_0^\ast \log\bar\mu^2 = 0\,.
\end{align}
In this way, we see that the $\log\bar\mu^2$ terms in \eqref{Wmunu} do not contribute to the spectral rate and a fortiori to the energy emission spectrum.

One then considers a detector with four-velocity $t^\mu$ placed at a spatial distance $r$ from the scattering event in the angular direction characterized by the null vector $\hat n^\mu$, so that
\begin{equation}\label{}
	x^\mu = u\, t^\mu + r\, \hat n^\nu\,,\qquad \hat n\cdot t=-1\,,
\end{equation}
and takes the asymptotic limit
\begin{equation}\label{}
	r\to\infty\,,\qquad
	u,\hat n^\mu\text{ fixed}.
\end{equation}
In this limit, the asymptotic field \eqref{hmunuWmunu} takes the form \cite{Weinberg:1995mt,Cristofoli:2021vyo,Donnay:2022ijr}
\begin{equation}\label{waveform}
	h_{\mu\nu}(x)
	\sim
	\int_0^\infty \frac{\lambda^{-\epsilon}}{(ir)^{1-\epsilon}}
	e^{-i\lambda u}
	\left(i
	 \tilde W_{\mu\nu} \big|_{k= \lambda \hat n} 
	 \right)
	\frac{d\lambda}{2(2\pi)^{2-\epsilon}}
	+
	(\text{c.c.})\,,
\end{equation}
so that using \eqref{Wmunu}
\begin{equation}\label{}
	h^{\mu\nu}(x)
	\sim
	\int_0^\infty \frac{\lambda^{-\epsilon}}{(ir)^{1-\epsilon}}
	e^{-i\lambda \left(u-\frac{G}{\epsilon}\,(m_1 u_1+m_2u_2)\cdot\hat n\right)}\,i
	\left[
	\tilde{\mathcal{B}}^{\mu\nu}_0
	+
	\tilde{\mathcal{B}}^{\mu\nu}_1
	+
	i
	\tilde{\mathcal{M}}^{\mu\nu}_1
	\right]_{k= \lambda \hat n} 
	\frac{d\lambda}{2(2\pi)^{2-\epsilon}}
	+
	(\text{c.c.}).
\end{equation}
In this way, the classical information extracted from the one-loop amplitude can be used to build the $\mathcal O(G^3)$ corrections to the asymptotic metric fluctuation $\sqrt{32\pi G}\,h_{\mu\nu}$, and its IR-divergent phase can be formally reabsorbed via a constant shift of the detector's retarded time \cite{Goldberger:2009qd,Porto:2012as}.

\section{Conclusions and Outlook}
\label{sec:outlook}

In this paper we calculated the $2\to3$ amplitude for the collision of two massive scalars and the emission of a graviton.
We focused on the near-forward regime, where the exchanged momenta are small, $\mathcal O(\hbar)$, compared to the masses, and on the soft region in which the loop momentum associated to the exchanged gravitons is of the same order. This allowed us to perform the integration  of the integrand first obtained in Ref.~\cite{Carrasco:2021bmu}, calculating the result up to and including $\mathcal O(\hbar^{-1})$ and $\mathcal O(\epsilon^0)$.
The result passes nontrivial consistency checks. It displays the appropriate structure of IR divergences predicted by Ref.~\cite{Weinberg:1965nx} as well as the correct factorization in the soft limit \cite{Weinberg:1964ew}.
After checking that the operator version of the eikonal exponentiation \cite{Damgaard:2021ipf,Cristofoli:2021jas,DiVecchia:2022piu} indeed works as expected and produces a classical, real and finite matrix element for the ``eikonal'', or more precisely $N$-operator \cite{Damgaard:2021ipf}, we sketched the calculation of the asymptotic waveform and spectral emission rates.
We derived an expression for such quantities, showed that the spectra are free of ambiguities, while the waveform itself is affected by an IR divergent phase or, once such an irrelevant phase is discarded, by the presence of the logarithm of an arbitrary scale.
 
The appearance of logarithms of arbitrary parameters  left behind by infrared divergences in waveform calculations is a manifestation of the so-called ``hereditary'' or ``tail'' effects \cite{Blanchet:1992br,Blanchet:1993ec,Blanchet:2013haa,Bern:2021dqo,Dlapa:2021npj} and can be ascribed to an arbitrariness in fixing the origin of the detector's retarded time \cite{Goldberger:2009qd,Porto:2012as}. It is interesting that these features already appear in the one-loop five-point calculation performed here, whereas they only intervene at three loops in the four-point calculation \cite{Bern:2021dqo,Liu:2021zxr,Bern:2021yeh,Dlapa:2021vgp}. We leave further investigations of this point for future work. 
Another interesting issue to which we plan to return is the comparison with subleading log-corrected soft theorems \cite{Sahoo:2018lxl,Saha:2019tub,Sahoo:2021ctw}, which are also intimately related to tail effects and to the long-range nature of the gravitational force in four spacetime dimensions. In analogy with the tree-level case, such checks will likely require to first obtain sufficient analytic control of the $b$-space expression of the waveform. More generally, but also in connection with the issue of infrared divergences, which here we removed by following the exponentiation \cite{Weinberg:1965nx}, it will be interesting to investigate how our results fit within the broader program of the eikonal operator and to understand whether an improved operator formalism is actually able to directly provide an infrared finite answer, possibly fixing the associated scale ambiguity.
Of course, for all such open issues, extremely valuable guidance will come from  comparisons with the available PN results (see e.g.~Ref.~\cite{Bini:2022enm} and references therein).

In the spirit of reverse-unitarity applications for classical gravitational scattering \cite{Herrmann:2021lqe,Herrmann:2021tct,DiVecchia:2021bdo,Riva:2021vnj,Mougiakakos:2022sic,Riva:2022fru,DiVecchia:2022piu,Heissenberg:2022tsn}, our result can be useful for verifying and extending calculations of radiative observables to $\mathcal O(G^4)$, including emitted energy-momentum \cite{Bini:2022enm,Dlapa:2022lmu} and angular momentum (see Refs.~\cite{Herrmann:2021lqe,Herrmann:2021tct,Manohar:2022dea,DiVecchia:2022piu} for the analogous $\mathcal O(G^3)$ results).
In this work we focused on the $\omega>0$ portion of the graviton spectrum, although of course interesting phenomena are associated with static effects \cite{Damour:2020tta,Mougiakakos:2021ckm,Riva:2021vnj,DiVecchia:2022owy,DiVecchia:2022piu} and require taking into account terms localized at $\omega=0$. Such additional contributions can be typically included by means of suitable dressed states, and are likely to be important in order to correctly account for angular momentum losses.

\subsection*{Acknowledgements}
We would like to thank Francesco Alessio, Paolo Di Vecchia, Kays Haddad, Martijn Hidding, Henrik Johansson, Gregor Kälin, Stephen Naculich, Ben Page, Rodolfo Russo, Augusto Sagnotti, Fei Teng, and Gabriele Veneziano for very useful discussions. The research of CH is supported by the Knut and Alice Wallenberg Foundation under grant KAW 2018.0116.
IVH is supported by the Knut and Alice Wallenberg Foundation under grants KAW
2018.0116 and KAW 2018.0162. Nordita is
partially supported by Nordforsk.

\appendix

\section{More on the Kinematics and on the Polarization Tensor}
\label{appendix:Kinematics}

It this appendix, we complement the material presented in Sections~\ref{ssec:physicalvariables} and \ref{ssec:polarization} concerning the properties of the kinematic variables and of the polarization tensor employed in the text.
Introducing the dual vectors
\begin{equation}\label{}
	\check u_1^\mu= \frac{y u_2^\mu-u_1^\mu}{y^2-1}\,,\qquad
	\check u_2^\mu= \frac{y u_1^\mu-u_2^\mu}{y^2-1}\,,
\end{equation}
so that $u_i\cdot \check u_j=-\delta_{ij}$ for $i,j=1,2$, we can decompose
\begin{equation}\label{}
	k^\mu= \omega_1 \check u_1^\mu + \omega_2 \check u_2^\mu +  k_\perp^\mu
\end{equation}
with $k_\perp \cdot u_i=0$ and similarly
\begin{equation}\label{q12decomp}
	q_1^\mu= -\omega_2 \check u_2^\mu + q_{1\perp}^\mu\,,\qquad
	q_2^\mu= -\omega_1 \check u_1^\mu + q_{2\perp}^\mu\,,\qquad
	q_{1\perp}^\mu + 	q_{2\perp}^\mu+ 	k_{\perp}^\mu=0
\end{equation}
where we used \eqref{orth5}.
The condition $k^2=0$ then takes the following form
\begin{equation}\label{kperp2}
	k_\perp^2 = \frac{-\omega_1^2+2y\omega_1\omega_2-\omega_2^2}{y^2-1} \ge 0\,,
\end{equation}
while $q_1^2$ and $q_2^2$ read
\begin{equation}\label{q1q2q1perpq2perp}
	q_1^2 = \frac{\omega_2^2}{y^2-1} + q_{1\perp}^2 \ge q_{1\perp}^2 \ge 0\,,\qquad
	q_2^2 = \frac{\omega_1^2}{y^2-1} + q_{2\perp}^2 \ge q_{2\perp}^2 \ge 0\,.
\end{equation}
The relations \eqref{kperp2}, \eqref{q1q2q1perpq2perp} imply that
\begin{equation}\label{}
	\frac{1}{y+\sqrt{y^2-1}}\le\frac{\omega_1}{\omega_2}\le y+\sqrt{y^2-1}\,,
	\qquad
	q_1^2 \ge \frac{\omega_2^2}{y^2-1}\,,\qquad q_2^2 \ge \frac{\omega_1^2}{y^2-1}\,.
\end{equation}
In addition, the Schwarz inequality $(q_{1\perp}\cdot q_{2\perp})^2\le q_{1\perp}^2 q_{2\perp}^2$ is equivalent to
\begin{equation}\label{Schwarz}
	\mathcal S = (y^2-1) (q_1^2-q_2^2)^2-4 y \omega_1 \omega_2 (q_1^2+q_2^2)+4 \omega_1^2q_1^2+4 \omega_2^2 q_2^2+4  \omega_1^2\omega_2^2 \le 0\,.
\end{equation}

The vector \eqref{polarizdecomp} can be rewritten as follows, after imposing the transversality condition \eqref{physpol},
\begin{equation}\label{}
	\varepsilon^{\mu} = c_1 \xi_1^\mu + c_2 \xi_2^\mu + d_+ (q_1^\mu + q_2^\mu)
\end{equation}
in terms of the two transverse vectors
\begin{equation}\label{}
	\xi_{1}^\mu = u_1^\mu - \omega_1\,\frac{q_1^\mu-q_2^\mu}{q_1^2-q_2^2}\,,\qquad
	\xi_{2}^\mu = u_2^\mu - \omega_2\,\frac{q_1^\mu-q_2^\mu}{q_1^2-q_2^2}\,.
\end{equation}
The vectors $k^\mu$, $\xi_1^\mu$, $\xi_2^\mu$ form a basis of the space of vectors $\xi^\mu$ such that $k\cdot \xi = 0$. All such vectors, except for those aligned with $k^\mu$, are spacelike, as can be easily seen by going to a frame where $k^\mu = (\kappa,0,0,\kappa)$, where $\xi^t=\xi^z$ and therefore $\xi^2 = (\xi^x)^2+(\xi^y)^2\ge 0$.
We can thus introduce 
\begin{equation}\label{}
	|\xi_1| = \sqrt{\xi_1^2} \ge 0\,,\qquad
	|\xi_1| = \sqrt{\xi_2^2} \ge 0\,,\qquad
	\Delta = 1 - \frac{(\xi_1\cdot \xi_2)^2}{\xi_1^2\xi_2^2} \ge 0\,,
\end{equation}
where the very last relation is the standard Cauchy--Schwarz inequality. Explicitly,  
\begin{equation}\label{}
	\xi_1^2 = -1 + \frac{4 \omega_1^2 q_1^2}{(q_1^2-q_2^2)^2}\,,\quad
	\xi_1\cdot \xi_2 = -y + \frac{2 \omega_1\omega_2}{(q_1^2-q_2^2)^2}\,(q_1^2+q_2^2)\,,\quad
	\xi_2^2 = -1 + \frac{4 \omega_2^2 q_2^2}{(q_1^2-q_2^2)^2}
\end{equation}
and (cf.~Eq.~\eqref{Schwarz})
\begin{equation}\label{}
	-\Delta\,\xi_1^2 \xi_2^2= 
	y^2-1 + \frac{4(\omega_1^2 q_1^2+\omega_2^2 q_2^2)-4 y \omega_1\omega_2(q_1^2+q_2^2)+4 \omega_1^2\omega_2^2}{(q_1^2-q_2^2)^2}\le 0\,.
\end{equation}

The polarization tensor \eqref{ourpol} can be made traceless by imposing
\begin{equation}\label{}
	\varepsilon_\mu \varepsilon^\mu = \xi_1^2 c_1^2  + 2 (\xi_2\cdot \xi_2)c_1 c_2  +  \xi_2^2 c_2^2= 0\,,
\end{equation}
which we can solve by allowing $c_1$ and $c_2$ to take complex values and letting
\begin{equation}\label{trace=0}
	|\xi_1|\,c_1 = \left[-\frac{(\xi_1\cdot \xi_2)}{|\xi_1|\,|\xi_2|} + i \sqrt{\Delta}\right]\,|\xi_2|\,c_2\,,
\end{equation}
or equivalently
\begin{equation}\label{}
	|\xi_1|\,c_1 = i\, e^{i\varphi_{12}}\,|\xi_2|\,c_2\,,\qquad
	\varphi_{12} = \arcsin\frac{(\xi_1\cdot \xi_2)}{|\xi_1|\,|\xi_2|}\,.
\end{equation}
One can identify
\begin{equation}\label{}
	\varepsilon^\mu = \frac{1}{\sqrt 2}\left(\varepsilon_1^\mu+ i \varepsilon_2^\nu\right)\,,
\end{equation}
in terms of real orthonormal vectors $\varepsilon_1^\mu$, $\varepsilon_2^\mu$,
\begin{equation}\label{}
	\varepsilon_i\cdot \varepsilon_j=\delta_{ij}\,,\qquad i,j=1,2\,,
\end{equation} 
after imposing the normalization condition 
\begin{equation}\label{normalization=1}
	\varepsilon^\ast \cdot \varepsilon 
	=
	\xi_1^2
	|c_1|^2 
	+
	(\xi_1 \cdot \xi_2)
	(c_1^\ast c_2 + c_2^\ast c_1)  
	+
	\xi_2^2
	|c_2|^2 
	=
	1\,,
\end{equation}
that is,
\begin{equation}\label{}
	|\xi_1|\ |c_1| = 1/\sqrt{2\Delta} = |\xi_2|\ |c_2|\,.
\end{equation}
Building the standard real polarization tensors,
\begin{equation}\label{}
	\varepsilon_+^{\mu\nu} = \frac{\varepsilon_1^\mu\varepsilon_1^\nu-\varepsilon_2^\mu\varepsilon_2^\nu}{\sqrt2}
	\,,\qquad
	\varepsilon_\times^{\mu\nu} = \frac{\varepsilon_1^\mu\varepsilon_2^\nu+\varepsilon_1^\nu\varepsilon_2^\mu}{\sqrt2}\,,
\end{equation}
the identification is
\begin{equation}\label{}
	\sqrt{2}\, 
	\operatorname{Re}\varepsilon^{\mu\nu} = \varepsilon_+^{\mu\nu}\,,\qquad
	\sqrt{2}\,
	\operatorname{Im}\varepsilon^{\mu\nu} =  \varepsilon_\times^{\mu\nu}\,.
\end{equation}
In the main body of the text, we mostly work without explicitly imposing the trace constraint \eqref{trace=0} and the normalization conditions \eqref{normalization=1}, treating $c_1$ and $c_2$ as formally independent.
In order to obtain $\mathcal A_1^{\mu\nu}$ from \eqref{contrA1}, one ought to first impose \eqref{trace=0}, \eqref{normalization=1} and then build
\begin{equation}\label{}
	\mathcal A_1^{\mu\nu} = \varepsilon^\mu (\varepsilon^\ast_\alpha\mathcal A_1^{\alpha\beta}\varepsilon_\beta^\ast)\varepsilon^\nu
	+
	\varepsilon^{\ast\mu} (\varepsilon_\alpha\mathcal A_1^{\alpha\beta}\varepsilon_\beta)\varepsilon^{\ast\nu}\,.
\end{equation} 

\section{Master Integrals}
\label{appendix:Masters}
In this appendix, we present the master integrals that we have used in order to perform the integration of the one-loop amplitude presented in the main text.
As is clear from the drawings in Table~\ref{16MIt}, it is enough to provide the expressions for 9 of them, since the remaining 7 are obtained by interchanging all labels $1$ and $2$, i.e.~applying the permutation $\sigma_4$.
\begin{equation}
	I_{0,0,1,1,0}=	\begin{tikzpicture}[baseline={([yshift=-0.5ex]current bounding box.center)},scale=1,node/.style={draw,shape=circle,fill=black,scale=0.4}]
		\path [draw, ultra thick, green!60!black] (-.65,0)--(.65,0);
		\path [draw, ultra thick, blue] (-.65,1)--(.65,1);
		\draw (0,.97) to[out=-60, in=60] (0,.03);
		\draw (0,.97) to[out=240, in=120] (0,.03);
		\draw (0,.97)--(.7,.5);
	\end{tikzpicture}= \frac{1}{\eps} +2 - 2 \Log{q_2}+\frac{\eps}{12}\left(48-\pi^2-48 \Log{q_2}+24 \Log{q_2}^2 \right) +\mathcal O(\epsilon)
\end{equation}
\vspace{3mm}
\begin{align}				  
& I_{1,0,0,1,0}
=	\begin{tikzpicture}[baseline={([yshift=-0.5ex]current bounding box.center)},scale=1,node/.style={draw,shape=circle,fill=black,scale=0.4}]
		\path [draw, ultra thick, green!60!black] (-.65,0)--(-.5,.5)--(.65,0);
		\path [draw, ultra thick, blue] (-.65,1)--(-.5,.5)--(.65,1);
		\path [draw,thin] (.5,.95)--(.7,.5);
		\draw (-.5,.5) to[out=-15, in=240] (.5,.95); 
	\end{tikzpicture}
= \frac{1}{\eps}\frac{q_2(w_{1E}^2-1)}{2 w_{1E}} \\
&+ \frac{q_2 (w_{1E}^2-1)}{w_{1E}} (\Log{w_{1E}-1}-1 -\Log{w_{1E}} +\Log{1 + w_{1E}} +\Log{q_2})  \nonumber \\
&+ \eps \frac{q_2 (w_{1E}^2-1)}{24 w_{1E}} (48 + 5 \pi^2+ 24(\Log{w_{1E}-1} -2 -\Log{w_{1E}} +\Log{1 + w_{1E}} +\Log{q_2}) \nonumber \\ 
& \times (\Log{-1 +w_{1E}} -\Log{w_{1E}} +\Log{1 + w_{1E}} + \Log{q_2})) +\mathcal O(\epsilon^2)
\nonumber
\end{align}
\vspace{3mm}
\begin{equation}
	I_{0,1,1,1,0}
	=				  
	\begin{tikzpicture}[baseline={([yshift=-0.5ex]current bounding box.center)},scale=1,node/.style={draw,shape=circle,fill=black,scale=0.4}]
		\path [draw, ultra thick, green!60!black] (-.65,0)--(.65,0);
		\path [draw, ultra thick, blue] (-.65,1)--(.65,1);
		\path [draw,thin] (0,.97)--(.5,.03);
		\path [draw,thin] (0,.97)--(-.5,.03);
		\path [draw,thin] (0,.97)--(.7,.5);
	\end{tikzpicture}= \frac{\pi^2}{2 q_2} +\mathcal O(\epsilon)
\end{equation}
The integrals $I_{0,0,1,1,0}$, $I_{1,0,0,1,0}$ and $I_{0,1,1,1,0}$ can be in fact evaluated in generic $D=4-2\epsilon$ with elementary methods. However, we opted to present their expansion for small $\epsilon$ in the form which is ready-to-use for the analytic continuation discussed in Section~\ref{ssec:euclidean}.
\begin{equation}	
	I_{1,0,1,1,0}
	=			  
	\begin{tikzpicture}[baseline={([yshift=-0.5ex]current bounding box.center)},scale=1,node/.style={draw,shape=circle,fill=black,scale=0.4}]
		\path [draw, ultra thick, green!60!black] (-.65,0)--(.65,0);
		\path [draw, ultra thick, blue] (-.65,1)--(.65,1);
		\path [draw,thin] (.5,.97)--(0,.03);
		\path [draw,thin] (-.5,.97)--(0,.03);
		\path [draw,thin] (.5,.97)--(.7,.5);
	\end{tikzpicture}= \frac{2 w_{1 E} \left( \pi^2+6 \Li{\frac{-1}{w_{1E}}} +6 \Li{\frac{1}{w_{1E}}}+3 \Log{w_{1E}}^2 \right)}{3 q_2 (1 + w_{1E}^2)} 
+\mathcal O(\epsilon)
\end{equation}
\vspace{3mm}
\begin{align}	
	&I_{1,1,0,1,0}=			  
	\begin{tikzpicture}[baseline={([yshift=-0.5ex]current bounding box.center)},scale=1,node/.style={draw,shape=circle,fill=black,scale=0.4}]
		\path [draw, ultra thick, green!60!black] (-.65,0)--(-.5,.5)--(.65,0);
		\path [draw, ultra thick, blue] (-.65,1)--(-.5,.5)--(.65,1);
		\path [draw,thin] (.5,.95)--(.5,.037);
		\path [draw,thin] (.5,.95)--(.7,.5);
	\end{tikzpicture} = \frac{1}{\eps}\frac{x_E \Log{x_E}}{(x_E^2-1)^2} \\
&+ \frac{x_E}{6(x_E^2-1)^2} (\pi^2-12\Li{\frac{-1}{x_E}}-12\Li{\frac{1}{x_E}} -6 \Log{x_E}\left( 2 \Log{-1 + w_{1E}} \right. \nonumber \\
	& -\left. 2 \Log{w_{1E}} +2 \Log{1 + w_{1E}} - 2 \Log{-1 + x_{E}}+3\Log{ x_{E}}-\Log{1 + x_{E}} +2\Log{q_2}\right) )
	\nonumber\\
	&+\mathcal O(\epsilon)\nonumber
\end{align}
\vspace{3mm}
\begin{align}	
&	I_{1,0,1,1,1}
	=			  
	\begin{tikzpicture}[baseline={([yshift=-0.5ex]current bounding box.center)},scale=1,node/.style={draw,shape=circle,fill=black,scale=0.4}]
		\path [draw, ultra thick, green!60!black] (-.65,0)--(.65,0);
		\path [draw, ultra thick, blue] (-.65,1)--(.65,1);
		\path [draw,thin] (.5,.97)--(0,.03);
		\path [draw,thin] (-.5,.97)--(0,.03);
		\path [draw,thin] (.25,.5)--(.7,.5);
	\end{tikzpicture}=\frac{1}{\eps^2}\frac{ w_{2E}}{2 q_2 q_1^2 (-1 + w_{1E}^2)}\\
\nonumber
&+ \frac{1}{\eps}\frac{w_{1E}}{ q_2 q_1^2 (-1 + w_{1E}^2)} \left( -\Log{w_{1E}-1}+\Log{w_{1E}} -\Log{1+w_{1E}}+ \Log{q_2}-2 \Log{q_1} \right) \\
\nonumber
&+ \frac{ w_{1E} }{8 q_2 q_1^2 (-1 + w_{1E}^2)} \left(-\pi^2 + 8 \Log{-1 + w_{1E}}^2 +8 \left( \Log{w_{1E}} -\Log{1 + w_{1E}}\right)\right. \nonumber 
\\ &\left.
\times \left( \Log{w_{1E}} - \Log{1 + w_{1E}} + 2 \Log{q_2} - 4 \Log{q_1}\right) \right. \nonumber\\
&\left. -16 \Log{-1 + w_{1E}} \left(\Log{w_{1E}} -\Log{1 + w_{1E}}  \right.  \right. \nonumber \\&\left. \left.+\Log{q_2} -2 \Log{q_1}\right)+ 64\Log{q_1}^2 +8\Log{q_2} \left(-3 \Log{q_2} + 4 \Log{q_1^2 -q_2^2}\right) \right. \nonumber \\&\left.-32 \Log{q_1} \Log{q_2 (q_1^2 - q_2^2)} +16\Li{\frac{q_2^2}{q_1^2}} \right)  
+\mathcal O(\epsilon)
\nonumber
\end{align}
\vspace{3mm}
\begin{align}
	&I_{1,1,0,1,1}=				  
	\begin{tikzpicture}[baseline={([yshift=-0.5ex]current bounding box.center)},scale=1,node/.style={draw,shape=circle,fill=black,scale=0.4}]
		\path [draw, ultra thick, green!60!black] (-.65,0)--(-.5,.5)--(.65,0);
		\path [draw, ultra thick, blue] (-.65,1)--(-.5,.5)--(.65,1);
		\path [draw,thin] (.5,.95)--(.5,.035);
		\path [draw,thin] (.5,.5)--(.7,.5);
	\end{tikzpicture}= \frac{1}{\eps^2}\frac{w_{1E}w_{2E}}{q_1q_2 \left(-1 + w_{1E}^2) (-1 + w_{2E}^2\right)} \\
\nonumber
&- \frac{1}{\eps}\frac{w_{1E}w_{2E}}{q_1q_2 \left(-1 + w_{1E}^2) (-1 + w_{2E}^2\right)}\left(\Log{-1 + w_{1E}} -\Log{w_{1E}} \right. \nonumber \\&\left.+ \Log{1 + w_{1E}} + \Log{-1 + w_{2E}} - 
	\Log{w_{2E}} + \Log{1 + w_{2E}} +\Log{q_1} +\Log{q_2} \right) \nonumber \\& +\frac{w_{1E}w_{2E}}{12 q_1q_2 \left(-1 + w_{1E}^2) (-1 + w_{2E}^2\right)} \Big(- 7\pi^2-12 \Log{x_E}^2+24 (\log (q_1)+\log (w_{2E}-1)
	\nonumber 
	\\
	&-\log (w_{2E})+\log (w_{2E}+1)) (\log (q_2)+\log (w_{1E}-1)-\log (w_{1E})+\log (w_{1E}+1)) \Big)
	\nonumber\\
	&+\mathcal O(\epsilon) \nonumber
\end{align}
\vspace{3mm}
\begin{align}	
&I_{1,1,1,1,0}=			  
	\begin{tikzpicture}[baseline={([yshift=-0.5ex]current bounding box.center)},scale=1,node/.style={draw,shape=circle,fill=black,scale=0.4}]
		\path [draw, ultra thick, green!60!black] (-.65,0)--(.65,0);
		\path [draw, ultra thick, blue] (-.65,1)--(.65,1);
		\path [draw,thin] (.5,.97)--(.5,.03);
		\path [draw,thin] (-.5,.97)--(-.5,.03);
		\path [draw,thin] (.5,.97)--(.7,.5);
	\end{tikzpicture}= 
-\frac{x_E \log (x_E)}{q_2^2 \left(x_E^2-1\right) \epsilon }
\\ \nonumber
&+
\frac{x_E}{6 q_2^2 \left(x_E^2-1\right)}
\Big[
-12 \operatorname{Li}_2\left(\frac{1}{x_E}\right)-12 \operatorname{Li}_2\left(-\frac{1}{x_E}\right) \nonumber\\
&+6 \log (x_E) (2 (\log (q_2)+\log (x_E+1))-2 \log (w_{1E}-1)+2 \log (w_{1E}) \nonumber \\
&-2 \log (w_{1E}+1)+2 \log (x_E-1)-3 \log (x_E))+\pi ^2
\Big]+\mathcal O(\epsilon)\nonumber
\end{align}
\vspace{3mm}
\begin{equation}
	I_{1,1,1,1,1}
	=				  
	\begin{tikzpicture}[baseline={([yshift=-0.5ex]current bounding box.center)},scale=1,node/.style={draw,shape=circle,fill=black,scale=0.4}]
		\path [draw, ultra thick, green!60!black] (-.65,0)--(.65,0);
		\path [draw, ultra thick, blue] (-.65,1)--(.65,1);
		\path [draw,thin] (.5,.97)--(.5,.03);
		\path [draw,thin] (-.5,.97)--(-.5,.03);
		\path [draw,thin] (.5,.5)--(.7,.5);
	\end{tikzpicture}= \frac{c_{2}}{\epsilon^2}+ \frac{c_{1}}{\epsilon}+{c_{0}}+\mathcal O(\epsilon)\,.
\end{equation}
We have obtained $c_2$, $c_1$ and $c_0$ by means of dimension shifting identities (see e.g.~\cite{Lee:2012cn,Lee:2013mka}). These express the 6-dimensional pentagon $I_{1,1,1,1,1}^{6D}$, which is finite, as a linear combination of 4-dimensional pentagon and box integrals. Since $I_{1,1,1,1,1}^{6D}$ only involves objects of transcendental weight 3, it can only contribute to the $\mathcal O(\epsilon)$ part of $I_{1,1,1,1,1}$ in $D=4-2\epsilon$. Therefore, the ``ansatz coefficients'' $c_2$, $c_1$ and $c_0$ can be fixed in terms of the box integrals already provided in the previous equations.

\section{Tree-level amplitudes}
\label{appendix:treelevel}

In this appendix we collect the tree level amplitudes that are useful in order to perform various checks on the $2\to3$ one-loop amplitude calculated in the text. We start from the tree-level $2\to2$ amplitude $\mathcal A_0^{(4)}$ involving four massive states,
\begin{equation}\label{A04pt}
		\mathcal A_0^{(4)}
		=
		\begin{gathered}
		\begin{tikzpicture}
			\path [draw, ultra thick, blue] (-4,2)--(-2,2);
			\path [draw, ultra thick, green!60!black] (-4,1)--(-2,1);
			\filldraw[black!20!white, thick] (-3,1.5) ellipse (.5 and .8);
			\draw[thick] (-3,1.5) ellipse (.5 and .8);
		\end{tikzpicture}
	\end{gathered}
=
\frac{4 \pi  G \bar{m}_1^2 \bar{m}_2^2 \left(8 y^2 (\epsilon -1)+4\right)}{q^2 (\epsilon -1)}-\frac{4 \pi  G \left(\bar{m}_1^2+\bar{m}_2^2\right)}{\epsilon -1}+\frac{\pi  G q^2 (3-2 \epsilon )}{\epsilon -1}\,.
\end{equation}
The $2\to3$ tree level amplitude $\mathcal A_0^{\mu\nu}$ involving four massive states and a graviton can be written as follows as the sum of a piece obtained form the double copy minus a piece only including the dilaton exchanges,
\begin{equation}\label{A05pt}
	\mathcal A_{0}^{\mu\nu}
	=
	\begin{gathered}
		\begin{tikzpicture}
			\path [draw, ultra thick, blue] (-4,2)--(-2,2);
			\path [draw, ultra thick, green!60!black] (-4,1)--(-2,1);
			\path [draw, red] (-3,1.5)--(-2,1.5);
			\filldraw[black!20!white, thick] (-3,1.5) ellipse (.5 and .8);
			\draw[thick] (-3,1.5) ellipse (.5 and .8);
		\end{tikzpicture}
	\end{gathered}
=
\mathcal A_{\text{dc}}^{\mu\nu} - \mathcal A_{\text{dil}}^{\mu\nu}\,,
\end{equation}
where \cite{DiVecchia:2020ymx}
\begin{equation}\label{}
	\begin{aligned}
		\mathcal A_{\text{dc}}^{\mu\nu} & =2\left(8 \pi G_N\right)^{\frac{3}{2}}\left\{\left(p_4 p_2\right)\left(p_3 p_1\right)\left(\frac{p_4^\mu}{p_4 k}-\frac{p_3^\mu}{p_3 k}\right)\left(\frac{p_2^\nu}{p_2 k}-\frac{p_1^\nu}{p_1 k}\right)+4 q_1^2 q_2^2\right. \\
		& \times\left[\frac{q_1^\mu\left(p_1 p_2\right)-p_2^\mu\left(p_1 k\right)+p_1^\mu\left(p_2 k\right)}{q_1^2 q_2^2}-\frac{p_3^\mu}{2 p_3 k}\left(\frac{p_1 p_2}{q_1^2}+\frac{1}{2}\right)+\frac{p_4^\mu}{2 p_4 k}\left(\frac{p_1 p_2}{q_2^2}+\frac{1}{2}\right)\right] \\
		& \left.\times\left[\frac{q_1^\nu\left(p_4 p_3\right)-p_3^\nu\left(p_4 k\right)+p_4^\nu\left(p_3 k\right)}{q_1^2 q_2^2}+\frac{p_1^\nu}{2 p_1 k}\left(\frac{p_4 p_3}{q_2^2}+\frac{1}{2}\right)-\frac{p_2^\nu}{2 p_2 k}\left(\frac{p_4 p_3}{q_1^2}+\frac{1}{2}\right)\right]\right\}
	\end{aligned}
\end{equation}
and
\begin{equation}\label{}
	\begin{aligned}
	&\frac{\mathcal A_\text{dil}^{\mu\nu}}{\sqrt2\,(\pi G)^{3/2}}
	=
	\frac{4 \left(q_1^2-q_2^2\right) q_1^\mu q_1^\nu \left(4 \bar{m}_1^2-q_1^2\right) \left(q_2^2-4 \bar{m}_2^2\right)}{q_2^2 (\epsilon -1) \left(-4 \omega _1 \bar{m}_1+q_1^2-q_2^2\right) \left(4 \omega _1 \bar{m}_1+q_1^2-q_2^2\right)}\\
	&+
	\frac{4 \left(q_1^2-q_2^2\right) q_2^\mu q_2^\nu \left(q_1^2-4 \bar{m}_1^2\right) \left(q_2^2-4 \bar{m}_2^2\right)}{q_1^2 (\epsilon -1) \left(-4 \omega _2 \bar{m}_2+q_1^2-q_2^2\right) \left(4 \omega _2 \bar{m}_2+q_1^2-q_2^2\right)}\\
	&
	-\frac{32 \omega _2 \bar{m}_2^2 \left(q_1^2-4 \bar{m}_1^2\right) \left(4 \bar{m}_2^2-q_2^2\right) q_1^{(\mu}u_2^{\nu)}}{q_1^2 (\epsilon -1) \left(-4 \omega _2 \bar{m}_2+q_1^2-q_2^2\right) \left(4 \omega _2 \bar{m}_2+q_1^2-q_2^2\right)}\\
	&
	-\frac{32 \omega _1 \bar{m}_1^2 \left(4 \bar{m}_1^2-q_1^2\right) \left(q_2^2-4 \bar{m}_2^2\right) u_1^{(\mu}q_2^{\nu)}}{q_2^2 (\epsilon -1) \left(-4 \omega _1 \bar{m}_1+q_1^2-q_2^2\right) \left(4 \omega _1 \bar{m}_1+q_1^2-q_2^2\right)}\\
	&
	-\frac{16 \left(q_1^2-q_2^2\right) \bar{m}_1^2 u_1^\mu u_2^\nu \left(4 \bar{m}_1^2-q_1^2\right) \left(q_2^2-4 \bar{m}_2^2\right)}{q_2^2 (\epsilon -1) \left(-4 \omega _1 \bar{m}_1+q_1^2-q_2^2\right) \left(4 \omega _1 \bar{m}_1+q_1^2-q_2^2\right)}\\
	&
	+\frac{16 \left(q_1^2-q_2^2\right) \bar{m}_2^2 u_2^\mu u_2^\nu \left(q_1^2-4 \bar{m}_1^2\right) \left(4 \bar{m}_2^2-q_2^2\right)}{q_1^2 (\epsilon -1) \left(-4 \omega _2 \bar{m}_2+q_1^2-q_2^2\right) \left(4 \omega _2 \bar{m}_2+q_1^2-q_2^2\right)}\\
	&
	\frac{\left(q_1^2+q_2^2\right) \left(q_1^2-4 \bar{m}_1^2\right) \left(q_2^2-4 \bar{m}_2^2\right) \eta^{\mu\nu}}{q_1^2 q_2^2 (\epsilon -1)}
-\frac{q_1^{(\mu}q_2^{\nu)}}{q_1^2 q_2^2}
	\frac{2 \left(q_1^2-4 \bar{m}_1^2\right) \left(q_2^2-4 \bar{m}_2^2\right) P_{q_1q_2}}{
		 (\epsilon -1) Q_{q_1q_2}}
	\end{aligned}
\end{equation}
with $A^{(\mu}B^{\nu)} = A^\mu B^\nu + A^\nu B^\mu$,
\begin{align}\label{}
	P_{q_1q_2}
	&=
	2 q_1^4 \left(8 \omega _1^2 \bar{m}_1^2-8 \omega _2^2 \bar{m}_2^2+3 q_2^4\right)+\left(q_2^4-16 \omega _1^2 \bar{m}_1^2\right) \left(16 \omega _2^2 \bar{m}_2^2+q_2^4\right)+q_1^8-4 q_2^2 q_1^6-4 q_2^6 q_1^2\,,\\
	Q_{q_1q_2}
	&
	=
	\left(-4 \omega _1 \bar{m}_1+q_1^2-q_2^2\right) \left(4 \omega _1 \bar{m}_1+q_1^2-q_2^2\right) \left(-4 \omega _2 \bar{m}_2+q_1^2-q_2^2\right) \left(4 \omega _2 \bar{m}_2+q_1^2-q_2^2\right).
\end{align}
The last ingredient is the ``Compton'' amplitude for the scattering of a graviton and a massive particle \cite{KoemansCollado:2019ggb}
\begin{equation}\label{ComptonAmplitude}
	\mathcal{A}^C_{\rho \sigma, \alpha \beta}
	=
	\begin{gathered}
		\begin{tikzpicture}
			\path [draw, ultra thick, blue] (-4,2)--(-2,2);
			\path [draw, thin, red] (-4,1)--(-2,1);
			\filldraw[black!20!white, thick] (-3,1.5) ellipse (.5 and .8);
			\draw[thick] (-3,1.5) ellipse (.5 and .8);
			\node at (-4,2)[left]{$k_1$};
			\node at (-2,2)[right]{$k_2$};
			\node at (-4,1)[left]{$(\rho\sigma)\ r_{1}$};
			\node at (-2,1)[right]{$r_{2}\ (\alpha\beta)$};
		\end{tikzpicture}
	\end{gathered}
\end{equation}
which reads
\begin{equation}\label{}
	\begin{aligned}
		& \mathcal{A}^C_{\rho \sigma, \alpha \beta}=2 \kappa^2 \frac{r_{1} \cdot\left(k_1+r_{2}\right) r_{1} \cdot k_1}{r_{1} \cdot r_{2}} \\
		& \times\left[\frac{\left(k_1+r_{2}\right)^\rho k_1^\alpha}{r_{1} \cdot\left(k_1+r_2\right)}-\frac{\left(k_1+r_1\right)^\alpha k_1^\rho}{k_1 \cdot r_1}+\eta^{\rho \alpha}\right]\left[\frac{\left(k_1+r_2\right)^\sigma k_1^\beta}{r_1 \cdot\left(k_1+r_2\right)}-\frac{\left(k_1+r_1\right)^\beta k_1^\sigma}{k_1 \cdot r_1}+\eta^{\sigma \beta}\right].
	\end{aligned}
\end{equation}

Both $\mathcal A_{\text{dc}}^{\mu\nu}$, $\mathcal A_{\text{dil}}^{\mu\nu}$ and $\mathcal A_C^{\rho\sigma,\alpha\beta}$ are gauge invariant,
\begin{equation}\label{}
	\mathcal A_{\text{dc}}^{\mu\nu}k_\mu = 0\,,\qquad
	\mathcal A_{\text{dil}}^{\mu\nu}k_\mu= 0\,,\qquad
	\mathcal A_C^{\rho\sigma,\alpha\beta}r_{1\rho}=0=\mathcal A_C^{\rho\sigma,\alpha\beta}r_{2\alpha}=0
\end{equation}
and can be glued into cuts by replacing the transverse-traceless projector $\Pi^{\mu\nu,\rho\sigma}$ over intermediate graviton states via
\begin{equation}\label{}
	\Pi^{\mu\nu,\rho\sigma}\to
	\frac{1}{2}\left(
	\eta^{\mu\rho}\eta^{\nu\sigma}
	+
	\eta^{\mu\sigma}\eta^{\nu\rho}
	-
	\frac{1}{1-\epsilon}\,
	\eta^{\mu\nu}\eta^{\rho\sigma}
	\right).
\end{equation}
It is easy to see that, provided the gravitons have nonzero frequency,
\begin{equation}\label{}
	\mathcal A_{\rho\sigma,\alpha\beta}^C = \mathcal A_{\rho\sigma,\alpha\beta}^{C[0]} + \mathcal O(\hbar)
\end{equation}
in the limit \eqref{hbarscaling}. The tree-level $2\to2$ amplitude \eqref{A04pt} behaves as $\hbar^{-2}$ to leading order, and only receives corrections analytic in $q^2$,
\begin{equation}\label{analytic}
	\mathcal A_0^{(4)} = \mathcal A_0^{(4)[-2]} + (\text{analytic in }q^2)\,.
\end{equation}
These corrections become short-range terms the Fourier transform \eqref{FT4} and are thus completely irrelevant to our analysis.
The tree-level $2\to3$ amplitude \eqref{A05pt} also behaves as $\hbar^{-2}$ to leading order,
\begin{equation}\label{CLtree}
	\mathcal A_0^{\mu\nu} = \mathcal A_0^{\mu\nu[-2]} + \mathcal O(\hbar^0)
\end{equation}
and is free of $\hbar^{-1}$ corrections. The property \eqref{CLtree} holds thanks to the choice of variables \eqref{parq}, \eqref{barINV} discussed in Subsection~\ref{ssec:physicalvariables}. 
The leading order $\mathcal A_0^{\mu\nu[-2]}$ coincides with the one given in \cite{Luna:2017dtq,DiVecchia:2021bdo}
(see also \cite{Comberiati:2022cpm}).

\providecommand{\href}[2]{#2}\begingroup\raggedright\endgroup

\end{document}